\def\Msun{\ifmmode M_{\odot} \else $M_{\odot}$\fi}
\documentclass[oldversion]{aa}  
\usepackage{graphicx}
\usepackage{longtable,lscape}
\usepackage{txfonts}
\usepackage{natbib}
\begin{document}
   \title{AMAZE. I. The evolution of the mass--metallicity relation at z$>$3
             \thanks{Based on data obtained at the VLT through the ESO program
			 178.B-0838}}

   \subtitle{}

   \author{R.~Maiolino\inst{1}
          \and
		  T.~Nagao\inst{2}
		  \and
          A.~Grazian\inst{1}
		  \and
		  F.~Cocchia\inst{1}
		  \and
	      A.~Marconi\inst{3}
		  \and
		  F.~Mannucci\inst{4}
		  \and
		  A.~Cimatti\inst{5}
		  \and
		  A.~Pipino\inst{6}
		  \and
		  S.~Ballero\inst{7}
		  \and
		  F.~Calura\inst{7}
		  \and
		  C.~Chiappini\inst{10,11}
		  \and
		  A.~Fontana\inst{1}
		  \and
		  G.L.~Granato\inst{8}
		  \and
		  F.~Matteucci\inst{7}
		  \and
		  G.~Pastorini\inst{3}
		  \and
		  L.~Pentericci\inst{1}
		  \and
		  G.~Risaliti\inst{9}
		  \and
		  M.~Salvati\inst{9}
		  \and
		  L.~Silva\inst{10}
		  }

   \offprints{R. Maiolino}

   \institute{INAF - Osservatorio Astronomico di Roma, via di Frascati 33,
    00040 Monte Porzio Catone, Italy
         \and
	National Astronomical Observatory of Japan, 2-21-1 Osawa, 
	Mitaka, Tokyo 181-8588, Japan
         \and
Dipartimento di Astronomia, Universit\`a di Firenze, Largo E. Fermi 2, 50125 Firenze, Italy
		 \and
		INAF - Istituto di Radioastronomia, Largo E. Fermi 5, 50125 Firenze, Italy
		\and
		Dipartimento di Astronomia, Universit\`a di Bologna, via Ranzani 1, 40127, Bologna,
		Italy
		\and
		Astrophysics, University of Oxford, Keble Road, Oxford OX1 3RH, United Kingdom
		\and
		Dipartimento di Astronomia, Universit\`a di Trieste, via Tiepolo 11, 34131 Trieste, Italy
		\and
		INAF - Osservatorio Astronomico di Padova, Vicolo Osservatorio 5, 35122 Padova, Italy
		\and
		 INAF - Osservatorio Astrofisico di Arcetri, Largo E. Fermi 5, 50125 Firenze, Italy
		\and
	INAF - Osservatorio Astronomico di Trieste, Via Tiepolo 11, 34131 Trieste, Italy
		\and
	Geneva Observatory, Geneva University, 51 chemins des Mailletes,
	CH 1290, Sauverny, Switzerland
             }

   \date{Received ; accepted }

 
  \abstract
   {
We present initial results of an ESO-VLT large programme (AMAZE) aimed at
   determining the evolution of the mass-metallicity relation
   at z$>$3 by means of deep near-IR spectroscopy.
   Gas metallicities are measured, for an initial sample of nine
   star forming galaxies at z$\sim$3.5, by means of optical nebular lines
   redshifted into the near-IR. Stellar masses are accurately determined
   by using Spitzer-IRAC data, which sample the rest-frame near-IR stellar light
   in these distant galaxies.
   When compared with previous
   surveys, the mass-metallicity relation inferred at z$\sim$3.5
   shows an evolution much stronger than observed at lower redshifts.
   The evolution is prominent
   even in massive galaxies, indicating that z$\sim$3 is
   an epoch of major action in terms of star formation and metal enrichment also
   for massive systems.
   There are also indications that the
   metallicity evolution of low mass galaxies is stronger
   relative to high mass systems, an
   effect which can be considered
   the chemical version of the galaxy downsizing.
   The mass-metallicity relation observed at z$\sim$3.5 is difficult
   to reconcile with the predictions of some hierarchical evolutionary models.
   Such discrepancies suggest that at z$>$3 galaxies are assembled mostly with
   relatively un-evolved sub-units, i.e. small galaxies with low star formation efficiency.
   The bulk of the star formation and metallicity evolution probably occurs
   once small galaxies are already assembled into bigger systems.
   }


   \maketitle
%

\section{Introduction}
\label{sec_intro}

The connection between galaxy mass and metallicity has been known
for a long time, starting with the seminal work of \cite{lequeux79}.
Given the difficulty in obtaining reliable galaxy masses several authors have
resorted in using the (optical) luminosity. In particular, various works
have reported a clear correlation between blue luminosity and metallicity,
in the sense that more luminous galaxies are characterized by higher metallicities
\citep[e.g.][]{garnett87,skillman89,brodie91,zaritsky94}. A major step forward
has been recently achieved by \cite{tremonti04}, who used optical photometric and
spectroscopic data of a sample of $\sim$53,000 galaxies from the SDSS to determine
the mass-metallicity relation of local galaxies (z$\sim$0.1).
Their work clearly showed that the primary physical parameter driving the correlation
with the gas metallicity is the (stellar) mass of galaxies and not their luminosity.
While \cite{tremonti04} focused on the gas metallicity,
a similar relation by using the SDSS survey
was found by \cite{gallazzi06} for what concerns the stellar metallicity.

Various physical processes may be responsible for the mass-metallicity relation.
One possibility is that outflows, generated by starburst winds, eject metal-enriched
gas into the IGM preferentially out of low-mass galaxies
(due to the shallow gravitational potential well),
making their enrichment less effective than in massive systems \citep[e.g.][]{tremonti04,
delucia04,finlator08}.
An alternative scenario is that
low mass systems are still at an early evolutionary stage and have still to convert
most of their gas into stars, hence they are poorly metal-enriched relative massive galaxies (which
are instead already evolved). This is the
so-called ``galaxy downsizing'' scenario, supported by various observational
evidences \citep[e.g.][]{juneau05,feulner05,franceschini06,asari07,perez07}, where massive galaxies
formed most of their stars rapidly and at high redshift, while low mass systems are
characterized by a slower evolution, which extends to low redshift.
Finally \cite{koppen07} ascribes the mass-metallicity relation to variations of the IMF
high-mass cutoff in different star forming environments.

The relative role of these processes in shaping the mass-metallicity relation is debated.
It is likely that each of them contributes at
least to some extent, since observational evidences have been found for all of them.
Each of these factors (outflows/feedback, downsizing, IMF)
has profound implications on the evolution of galaxies. Therefore, it is clear
that the mass-metallicity relation contains a wealth of information useful
to constrain models of galaxy formation and evolution.
Indeed, any model of galaxy evolution is now required to
match the mass--metallicity relation observed locally
\citep[e.g.][]{kobayashi07,brooks07,derossi07,dave07,
dalcanton07,delucia04,tissera05,bouche06,bouche07,koppen07,cidfernandes07,finlator08,tassis08}.
However, different models predict
different evolutionary patterns of the mass-metallicity relation as a function
of redshift, and observational data are required to test and discriminate among
them.

Observational constraints of the mass-metallicity relation
have been obtained up to z$\sim$2.2 thanks to various deep surveys
\citep{savaglio05,liang06,erb06}. Additional observational studies
have investigated the evolution of the luminosity-metallicity relation
or, more generally, the metallicity of high-z star forming galaxies
\citep{kobulnicky00,kobulnicky03,kobulnicky04,maier04,maier05,maier06,forster06}.
Note that all of these studies
refer to the gas metallicity, while limited work has been done on the stellar metallicity
of high redshift sources
\citep{rix04,demello04,halliday08,mehlert06} due to difficulties in obtaining high S/N spectra
on the stellar continuum.
The general observational result is that the mass-metallicity relation (as well
as the luminosity-metallicity relation) evolves, in the sense that on average
higher redshift galaxies are characterized by lower metallicities (at a given mass).
Whether the relation evolves also in terms of its shape is still matter of debate.
Theoretical models can generally cope with the observed evolution, within both
the theoretical and observational uncertainties \citep[e.g.][]{kobayashi07,derossi07,brooks07,
finlator08}.

At z$\ge$3, except for a few individual studies, little work has been currently
done for what concerns the mass-metallicity relation. \cite{pettini01} have measured
the metallicity for a small sample of Lyman Break Galaxies (LBG) at z$\sim$3, but
without investigating the mass-metallicity relation. The metallicity evolution has
been investigated in DLA systems \citep{prochaska03,kulkarni05,akerman05}. However, a
study of the mass-metallicity relation for high-z absorption systems has not been
properly performed, due to difficulties in measuring the associated
stellar masses. The closest result is the finding of a relation between metallicity
and velocity dispersion (probably related to the mass) inferred by the width of the
absorption systems \citep{ledoux06,prochaska07}.

We have undertaken a large observing programme at ESO-VLT aimed at carefully
determining the mass-metallicity relation at z$>$3 for a sizeable sample of galaxies.
The final goal is to obtain a detailed description of the evolution of the
mass-metallicity evolution through the cosmic epochs, and therefore constrain
galaxy evolutionary scenarios.
In this paper we present preliminary results obtained by such a program, and discuss
the relevant implications for our understanding of the galaxy evolution at high
redshift. Throughout the paper we adopt the following cosmological parameters: $\rm
H_0=71~km~s^{-1}~Mpc^{-1}$, $\rm \Omega _{\Lambda}=0.73$ and
$\rm \Omega _m=0.27$ \citep{spergel03}.


\section{The AMAZE program} \label{sec_amaze}

AMAZE (Assessing the Mass-Abundance redshift[-Z] Evolution) is an ESO large program
aimed at determining the mass-metallicity relation in the redshift range 3$<$z$<$5.
Observations are being
performed with SINFONI \citep{eisenhauer03}, the near-IR integral field spectrometer
at VLT, for a total of 180 hours, distributed in three semesters. Observations are expected
to be completed in mid-2008. The target sample consists of about 30 Lyman Break Galaxies
(LBGs), most of which at 3$<$z$<$3.7, and only a few of them at
4.3$<$z$<$5.2. A more detailed discussion on the sample
selection is given in section~\ref{sec_samp}.
In this paper we present
preliminary results based on a first set of data for 9 galaxies,
and restricted to the redshift range 3$<$z$<$3.7 (Tab.~\ref{tab_samp}).
The integration times range from 3 to 7.5 hours on source.
The goal of the SINFONI observations is to determine
the gas metallicities by means of a combination of strong line diagnostics
based on H$\beta$ and [OIII]5007 shifted into the K band, as well as
[OII]3727 and [NeIII]3870 shifted into the H band for sources at 3$<$z$<$3.7.
At 4.3$<$z$<$5.2 we only rely on the  [OII]/[NeIII] ratio observed in the K band
(but sources in this redshift interval will not be discussed here).
Details on the gas metallicity determination are given in section~\ref{sec_met}.

The two-dimensional spectroscopic capabilities of SINFONI are obviously exploited also
to map the emission lines. However, the two-dimensional analysis of the spectra
goes beyond the scope of this paper and will be discussed in a separate paper.

\subsection{Sample selection} \label{sec_samp}

Galaxies in our sample are selected among z$>$3 galaxies primarily identified through the
Lyman--break technique, mostly from the \cite{steidel03} survey and from the deep
spectroscopic surveys in the Chandra Deep Field South (CDFS) \citep[e.g.][]{vanzella06},
but we also included some lensed galaxies \citep[e.g.][]{frye02,frye07}
to better explore the low mass end (but none of the lensed galaxies is in
the preliminary sample presented in this paper).
Galaxies were selected only amongst those with highly
reliable spectroscopic redshift \citep[e.g. flagged as ``A'' in ][]{vanzella06}. We
required that the redshift is such that the emission lines of interest for the metallicity
determination
([OIII],H$\beta$,[OII],[NeIII]) are out of strong sky emission lines and out of deep
atmospheric absorption features. Actually, these requirement could not always be
fulfilled for all of the emission lines (also because sometimes the
redshift determined through optical spectra is not accurate, due to winds affecting UV-rest frame
features, or IGM absorption of the Ly$\alpha$); however, in these cases
any contamination by OH lines or lower S/N due to atmospheric absorption
will be fully taken into account when estimating the metallicity.

The additional requirement is that the source has been observed with at least two of the
Spitzer-IRAC bands, which at these redshifts sample the rest-frame near-IR light. IRAC data are
required for a reliable determination of the stellar mass (see \S\ref{sec_mass}).
Finally, we excluded sources whose optical spectrum shows indications for the presence of
an AGN. Moreover, for most of the sources we also required that deep hard X-ray data and
mid-IR Spitzer-MIPS data are available, to better exclude the presence of a hidden AGN,
as discussed in \S\ref{sec_agn}.

In this paper we only present results for an initial sub-sample of nine sources
at 3$<$z$<$3.7, for which data have been already obtained and reduced. The list of sources,
along with some of their photometric properties, is given in Tab.~\ref{tab_samp}.
Tab.~\ref{tab_phys} lists some of the physical properties of these sources as inferred from their
broad-band spectral energy distribution. A detailed discussion on the extraction of these parameters
will be given in \S\ref{sec_mass}.

\subsection{AGNs removal} \label{sec_agn}

The presence of an AGN, contributing to the gas ionization, affects
the observed emission line ratios. In this case the metallicity diagnostic diagrams
calibrated on star forming galaxies are not usable, since the excitation mechanism
is totally different. As a consequence, galaxies hosting AGNs must be carefully avoided.

A first step is to exclude galaxies whose optical spectrum (UV rest frame) shows
indications for the presence of an AGN (e.g. NV, CIV, HeII, or broad Ly$\alpha$).
However, the absence of optical-UV AGN-like lines is a required condition,
but not sufficient to
rule out the presence of an AGN. Indeed, even if an AGN is present, the associated
optical-UV emission lines may be undetected either because obscured by dust (either on
small scales, for the BLR, or on larger scales, for the NLR), or because
their Narrow Line Region is not developed \citep[e.g.][]{maiolino03,martinez06}.

An additional constraint comes from hard X-ray data.
In many of the fields used by us, deep Chandra observations allow the detection of
obscured (Compton thin) AGNs up to z$\sim$4, even at Seyfert-like luminosities.
Therefore, an additional requirement was that our sources are not detected in the
hard X-rays (2--8~keV). They should not be detected also in the soft X-rays (0.5-2~keV) at
a level higher than expected by strongly star forming galaxies (actually none of the
galaxies is detected even in the soft band). In the CDFS the deep X-ray data allow us
to exclude the presence of obscured, Compton thin AGNs with 2--10~keV luminosity
higher than about $\rm 5\times 10^{43}~erg~s^{-1}$ (i.e. in the Seyfert range).

However, current X-ray surveys are not deep enough to detect Compton thick AGNs, whose
emission is strongly suppressed even in the hard X-rays. This issue was made clear by
recent Spitzer results
\citep{martinez07,alonso06,polletta06,fiore07,daddi07}. Indeed,
Spitzer observations have revealed the presence of obscured AGNs, through the associated
mid-IR hot dust emission, even in high-z galaxies that are undetected in deep hard X-ray
observations. These are shown to be high-z
Compton thick (or nearly Compton thick) AGNs, which remained
elusive to previous optical and X-ray surveys. Deep 24$\mu$m Spitzer-MIPS data were
found to
be particularly efficient to identify high-z obscured AGNs, even at relatively low
intrinsic luminosities. As a consequence, we requested that our sources have
deep MIPS data at 24$\mu$m.
\cite{fiore07} showed that the mid-IR excess relative to the optical emission
($\rm F_{24\mu m}/F_R$), is a good tracer of obscured AGNs at high redshift.
As listed in Tab~\ref{tab_samp}, most of our sources are undetected at 24$\mu$m.
More specifically, all our sources have a ratio $\rm F_{24\mu m}/F_R < 10$, which is 
significantly lower than expected for Sy2s and QSO2s at z$\sim$3 ($\rm F_{24\mu m}/F_R \sim 20-3~10^4$),
therefore ruling out the presence of AGNs (both Compton thick and thin)
even at Seyfert-like luminosities.

\section{Observations and data reduction} \label{sec_obs}

The near-IR spectroscopic observations were obtained by means of
SINFONI, the integral field spectrometer at VLT. SINFONI was used in its seeing-limited
mode, with the 0.25$''$ pixel scale and with the H+K grism, yielding a spectral resolution
R$\sim$1500 over the spectral range 1.45--2.41$\mu$m.

  \begin{figure*}
   \centering
   \includegraphics[width=12cm]{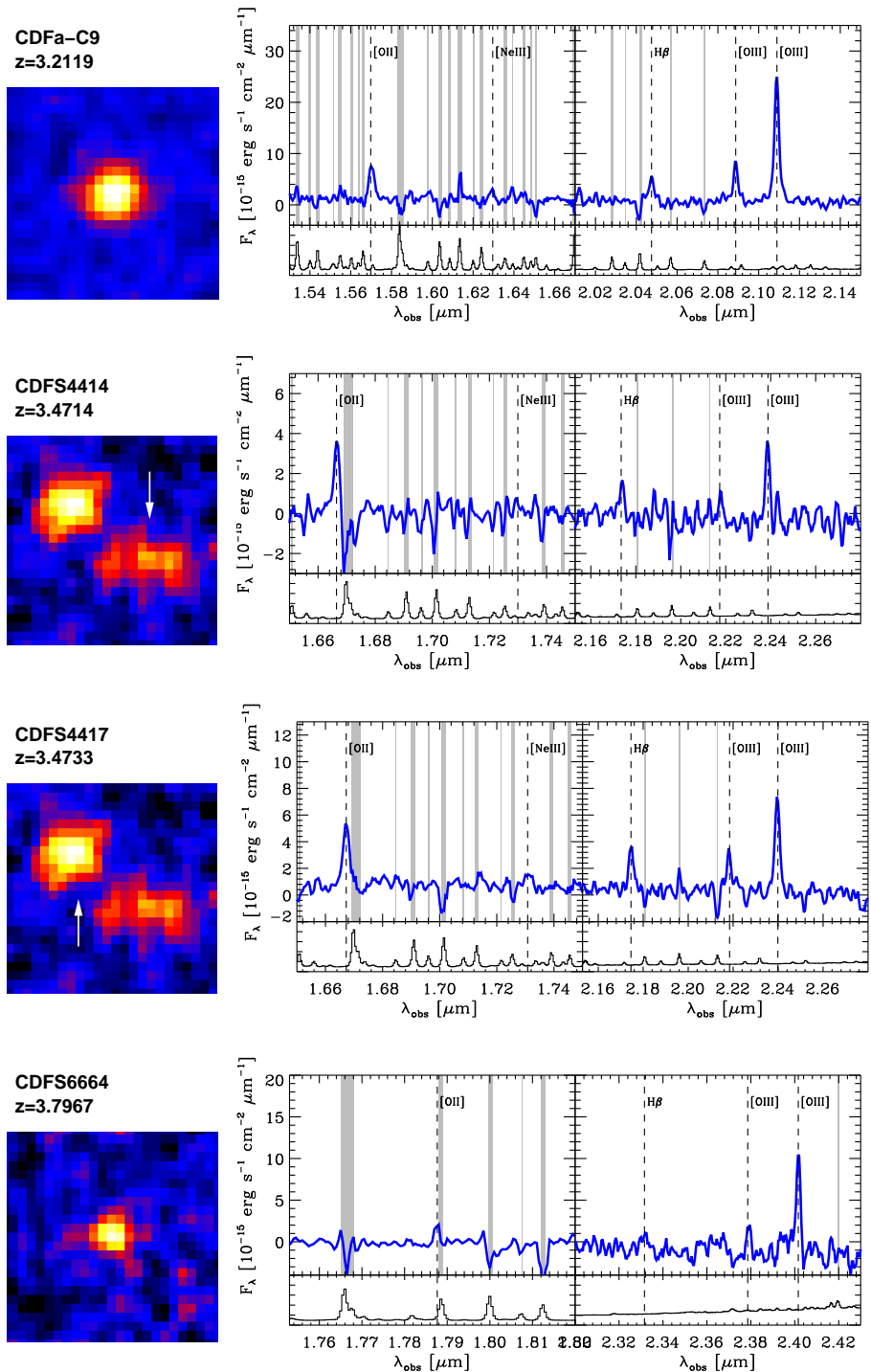}
\caption{[OIII]5007 line maps (left) and near-IR spectra (right)
of the galaxies in the AMAZE sample presented here. Each [OIII]5007 map has a size of
$\rm 3''\times 3''$.
The vertical dotted lines in the spectra
indicate the expected location of nebular emission lines.
The bottom panels show the sky spectrum.
The shaded vertical regions overlaid on each spectrum highlight
spectral region affected by strong sky emission lines.
}
  \label{fig_spec1}
\end{figure*}

  \begin{figure*}
   \centering
   \includegraphics[width=12cm]{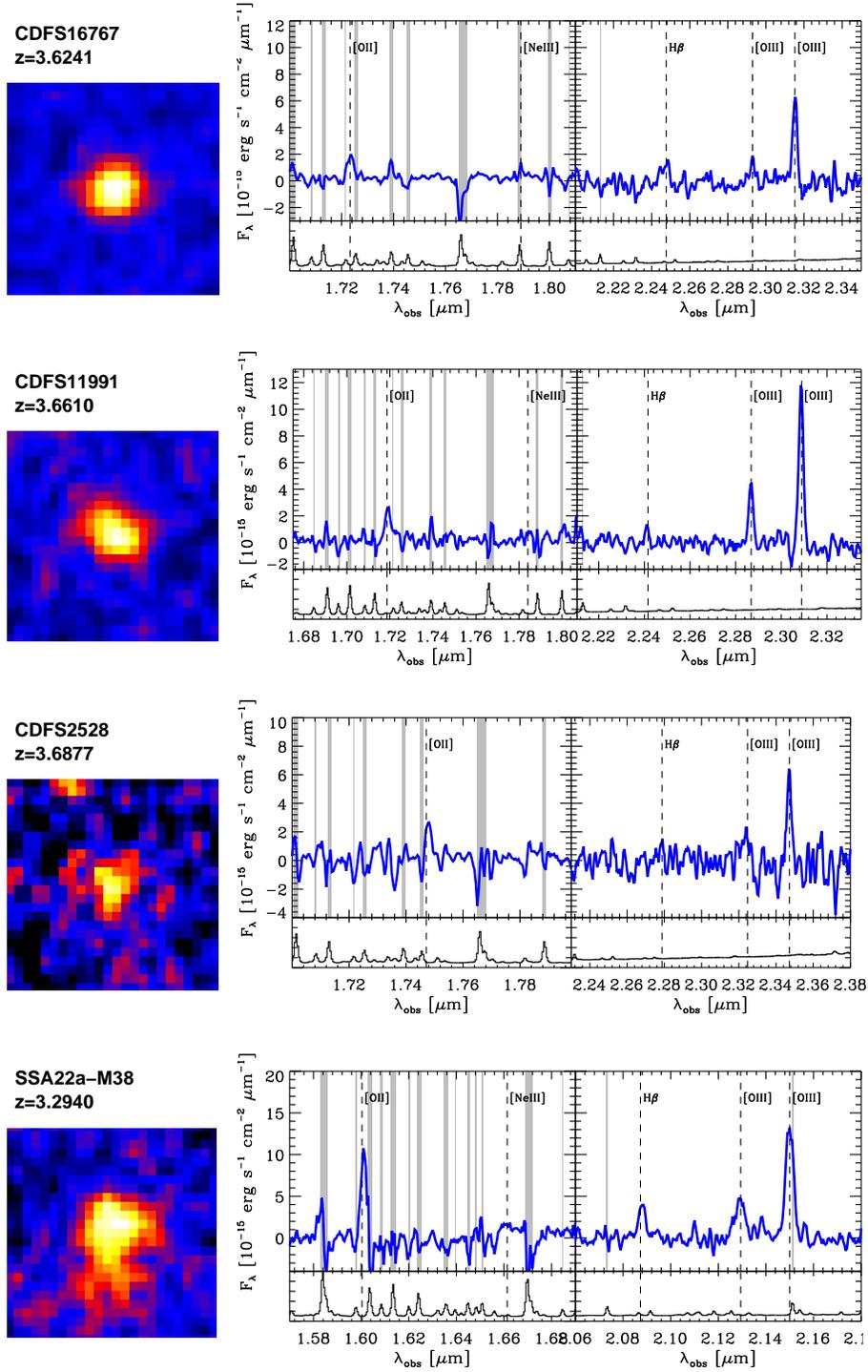}
\caption{Same as Fig.~\ref{fig_spec1} for four additional sources.
}
  \label{fig_spec2}
\end{figure*}

  \begin{figure*}
   \centering
   \includegraphics[width=12cm]{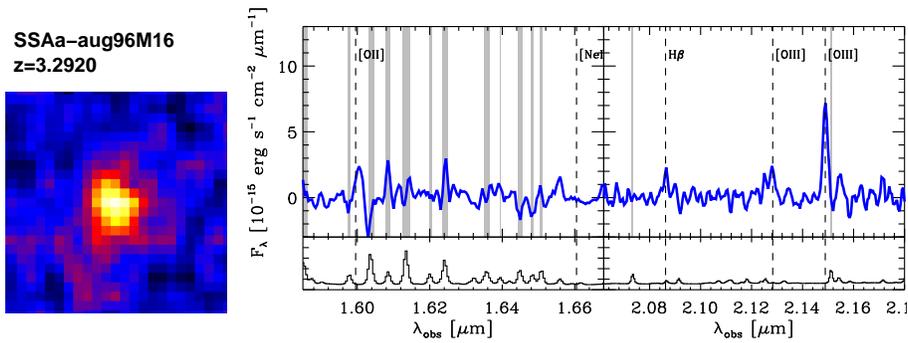}
\caption{Same as Figs.~\ref{fig_spec1}-\ref{fig_spec2} for one additional source.
}
  \label{fig_spec3}
\end{figure*}

  \begin{figure*}
   \centering
   \includegraphics[width=12cm]{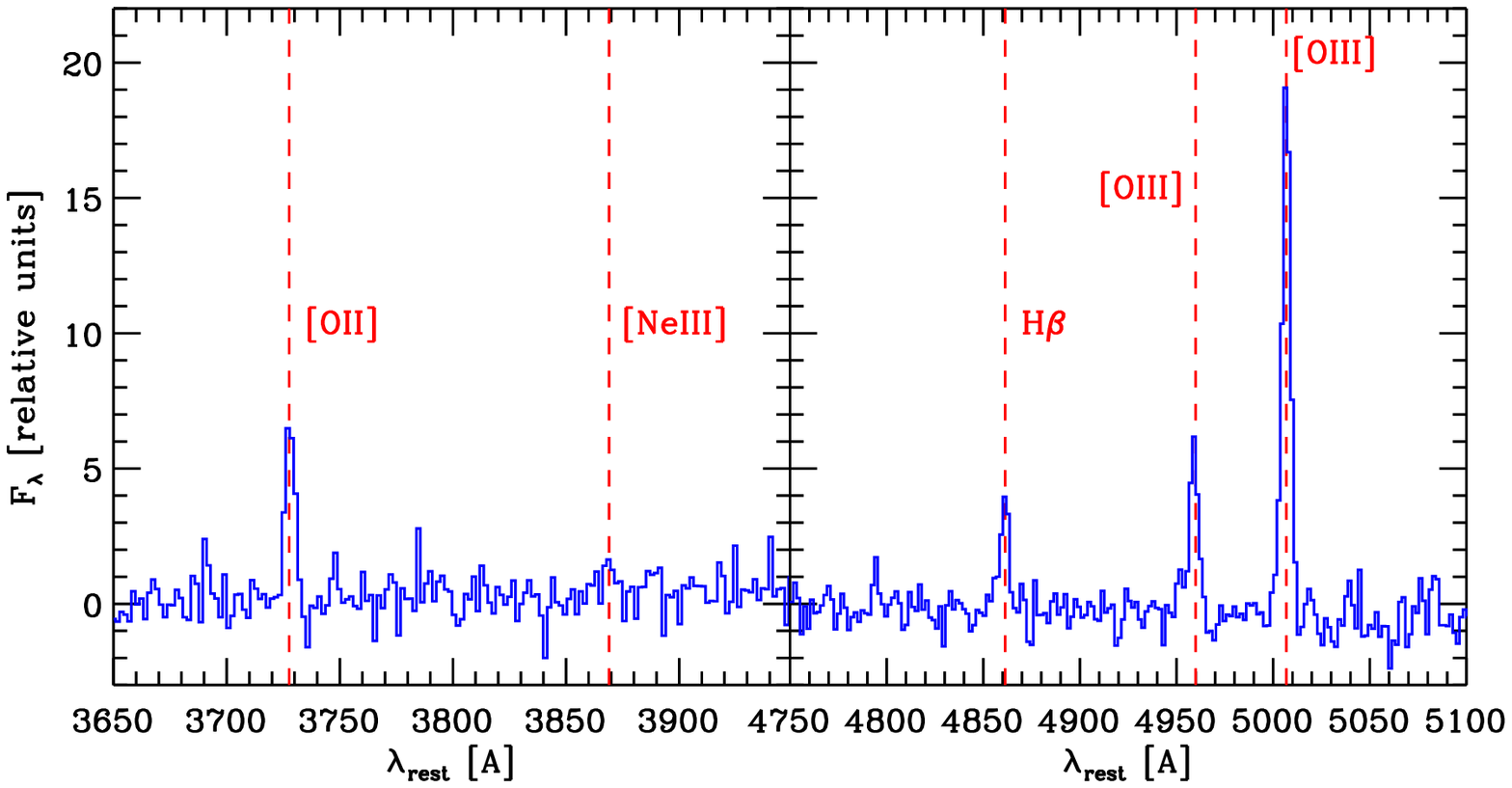}
\caption{Composite spectrum of the nine sources presented in this paper.
}
  \label{fig_stack}
\end{figure*}

Each target was acquired through a blind offset from a nearby bright star.
Each observing block consisted of 10 integrations, 5 minutes each, obtained by nodding the
position of the source within the $\rm 8''\times8''$ SINFONI field of view (generally by locating the
source in two opposite corners). This observational procedure allows background subtraction
by using frames contiguous in time, but with the source in different locations.
Moreover, the source was never located at the same position in the FOV: a minimal
dithering of 0.5$''$ was required, so that instrumental artifacts can be minimized when
the individual observations are aligned and combined together.
The (K-band) seeing during the observations was generally better than 0.8$''$.
Each source was observed with a number of observing blocks ranging
from 5 to 9. Some observing block was discarded because the seeing was much worse,
or the background much higher, with respect to the other observing blocks.
The total on-source integration times are listed in Tab.~\ref{tab_samp}.

Data were reduced by using the ESO-SINFONI pipeline (version 3.6.1). The pipeline
subtracts the sky from the temporally contiguous frames,
flat-fields the images, spectrally calibrates each individual
slice and then reconstructs the cube. Residual sky emission was accounted for by
removing the median of each spectral plane; this is feasible because our source occupy only
a small part of the field of view.
In some cases we performed an additional step in the background subtraction (which
resulted imperfect with the previous method probably because of minor uncertainties in the flat-fielding)
by sampling the
sky in a region outside the source (either annular or another region in the field of view
observed with the same effective integration) and rescaling it to optimally subtract the sky lines
on the spectrum of the source.
Individual cubes where aligned in the spatial
direction by relying on the telescope offsets and then averaging them together by applying
a 2$\sigma$ clipping to remove bad pixels and cosmic rays.

The atmospheric absorption and instrumental response were corrected by dividing the
spectrum of the scientific target by the spectrum
of a star (spectral type OV-BV or GV) taken close to the source, both in
time and in elevation. The intrinsic spectrum of the star was removed by dividing the
observed stellar spectrum by the appropriate template given in \cite{pickles98}, or by the
solar spectrum in the case of GV stars \citep{maiolino96}.

\section{Results and data analysis} \label{sec_res}

We extracted the spectra within a fixed aperture of 0.75$''$ in
diameter (corresponding to $\sim$6~kpc projected on sources at z$\sim$3.5),
which in most cases encloses more than 70\% of the emission line flux and generally
maximizes the S/N ratio. However, one should keep in mind that there are metallicity gradients
within each galaxy and therefore the aperture choice may introduce biases, especially
for what concerns the comparison with low redshift surveys. This issue
will be discussed more extensively in \S\ref{sec_apert}.
An exception is SSA22a-aug96M36, whose line emission extends significantly
beyond the 0.75$''$ aperture; in this case we adopted an aperture of 1.25$''$.
The resulting spectra, smoothed with a 2 pixel boxcar to improve the signal-to-noise, are
shown in Figs.~\ref{fig_spec1}-\ref{fig_spec3}. The location of [OII]3727, [NeIII]3870
(when observable), H$\beta$, [OIII]4959 and [OIII]5007 is indicated with vertical dashed lines.
In each spectrum the bottom panels show the sky spectrum. The shaded vertical regions
overlaid on each spectrum highlight spectral regions affected by strong sky emission lines.
In Figs.~\ref{fig_spec1}-\ref{fig_spec3} we also show the [OIII]5007 line map of each source.

Fig.~\ref{fig_stack} shows the composite spectrum of all nine sources, obtained by
shifting the spectra to the rest-frame, resampling them to a common wavelength scale,
normalizing them by the flux of H$\beta$ and averaging them. We excluded spectral
regions strongly affected by atmospheric absorption within individual spectra.

The stellar continuum is detected only in a few cases, and
even in these cases the continuum is only seen in the map produced by stacking the cube
in the spectral direction in the K or H band.

The emission line fluxes were measured by fitting a single gaussian over a linearly
interpolated, underlying continuum (which may be some weak stellar continuum or, more often,
residual thermal background or residual bias subtraction).
The resulting line fluxes are given
in Tab.~\ref{tab_met}. Note that for what H$\beta$ is concerned we do not
perform any subtraction of a stellar component, since the stellar continuum
is always very week and generally undetected, hence the correction for any putative stellar
H$\beta$ is negligible. Some authors apply a fixed correction of 2\AA \ 
for the EW of
a putative H$\beta$ in absorption; in our case such a correction would
generally affect the inferred
metallicities by less than 0.03 dex.

\section{The gas metallicity} \label{sec_met}

\subsection{Metallicity diagnostics and calibrations} \label{sec_met_cal}

The only method
to determine the gas metallicity in faint distant emission line
galaxies is to use strong
line metallicity diagnostics. Essentially, the ratio between various
strong, optical emission lines is found to depend on the gas metallicity,
either directly and/or through other
dependences (e.g. the metallicity dependence of the ionization parameter, gas density,
hardness of ionizing radiation, etc...).
Various strong line ratios have been calibrated against metallicity, either determined
``directly'' (e.g. through the electron temperature T$_e$ method) or ``indirectly''
(e.g. through photoionization models).
However, such calibrations have often been
performed in relatively narrow metallicity intervals, not adequate to explore the
wide metallicity range spanned by galaxies through the cosmic epochs, as we shall see.
Another serious problem is that such calibrations are often inconsistent with each other:
the same galaxy is found to have significantly different metallicities if different
strong-line diagnostics are adopted.
This issue has been reviewed in detail by
\cite{kewley08}. Obviously, a wrong intercalibration between
different metallicity diagnostics has dramatic implications for the investigation of the
metallicity evolution. Indeed, at different redshifts people have
observed different emission lines, depending on the adopted band, and therefore an incorrect
intercalibration between the various diagnostics may hamper the capability of
investigating evolutionary effects, or may even introduce artificial trends.

  \begin{figure*}[!t]
   \centering
   \includegraphics[width=12cm]{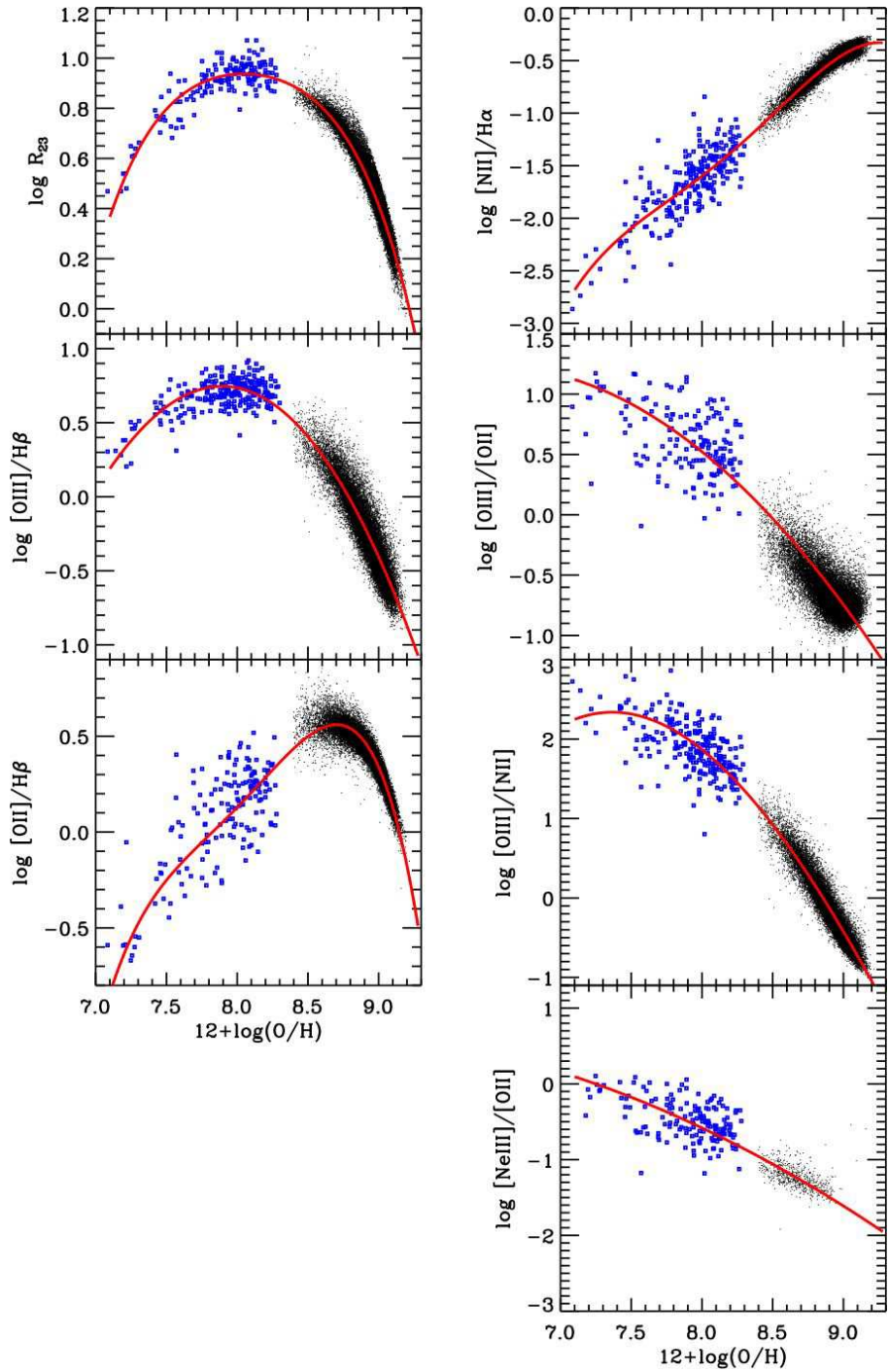}
\caption{Relations between strong emission line ratios and gas metallicity. Blue squares are
low metallicity galaxies \citep[from ][]{nagao06} for which the metallicity is inferred through
the electron temperature $\rm T_e$ method. Black dots are SDSS DR4 galaxies for which
metallicities are inferred through the photoionization models in \cite{kewley02}.
The red line curve shows a polynomial fit to the relations in the form given by
Eq.~\ref{eq_met_cal} with the coefficients given in Tab.~\ref{tab_met_cal}
}
  \label{fig_cal}
\end{figure*}

To minimize this issue it is recommended to use the same strong line calibration method
for all objects at various redshifts and from different surveys,
or to convert the strong line calibrations adopted by different authors to a common calibration
scale \citep[e.g. by using the conversion formulas given in][]{kewley08}. However,
the problem that we face in this paper is that no single strong line calibration method
exists over the wide metallicity range spanned by galaxies through the cosmic epochs (as we
shall see). Some methods nominally span a somewhat wider metallicity range, but they are
known to run into serious troubles in some metallicity intervals.

The electron temperature $\rm T_e$ method (e.g. by exploiting the intensity of
the [OIII]4636 auroral line) provides a good measure of the metallicity
below about 12+log(O/H)$<$8.3 \citep[e.g.][]{pilyugin01,pettini04}, and can be used to calibrate
the strong line ratios in this range. The reliability of the $\rm T_e$ method in the low
metallicity range is confirmed by the comparison with the stellar (OB) photospheric metallicity
measurements \citep{bresolin06b,bresolin07b}.
The $\rm T_e$ method has been extended to higher metallicities by various authors
\citep{kennicutt03,garnett04,liang07,yin07}.
However, at high metallicities
the $\rm T_e$ method tends to saturate and to underestimate significantly the true metallicity, due
to temperature fluctuations and gradients, both within individual HII regions and over
the whole galaxy.
This issue is expected theoretically \citep{stasinska05} and verified observationally
by the comparison with the metallicities determined through recombination lines,
which are insensitive to temperature fluctuations \citep{bresolin06a,bresolin07a}.

Photoionization models are an alternative way of calibrating strong line ratios
\citep[e.g.][]{tremonti04,kewley02,zaritsky94}, especially at high metallicities,
where most of these studies apply.
However, all photoionization models are subject to significant uncertainties and
possible systematic effects.
The observed spread in calibration between different  models highlights this problem
\citep{kewley08}. The photoionization models presented
in \cite{kewley02} are probably not free from the uncertainties and possible systematic
effects discussed above, however they provide results which are intermediate among
other photoionization models \citep{kewley08},
and therefore can be considered fairly representative of this class of calibrations.
Moreover, independent ``direct'' determinations of the metallicity
\citep[by exploiting the temperature--insensitive method
of recombination lines, ][]{bresolin06a,bresolin07a} are  in fair
agreement with the photoionization models provided by \cite{kewley02}.
The latter did not investigate
photoionization models at metallicities 12+log(O/H)$<$8.4. Other studies
attempt to extend photoionization models to 12+log(O/H)$<$8.3, but fail to reproduce
the observed line ratios \citep[e.g.][]{dopita06}.

  \begin{figure*}[!ht]
   \centering
   \includegraphics[width=15cm]{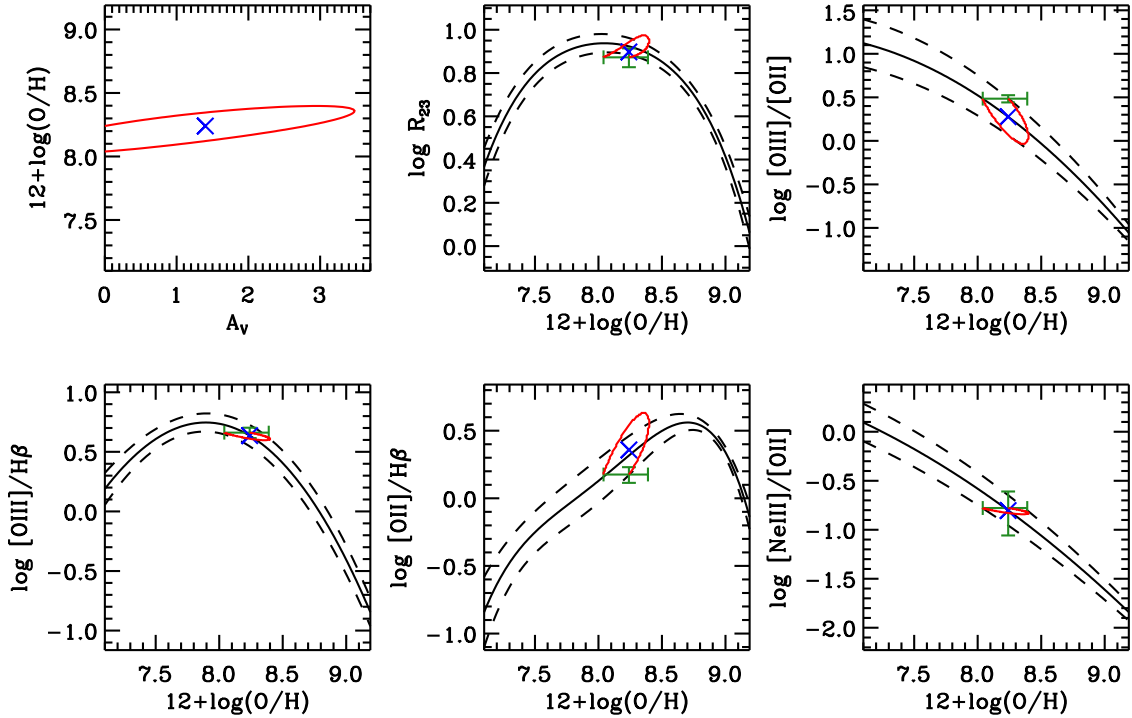}
\caption{Example of the diagnostic tools used to determine the metallicity
in the specific case of the composite spectrum. The upper left panel shows the
best solution (blue cross) and the 1$\sigma$ confidence level
in the $\rm A_V$--metallicity plane. In the other panels the black solid line
(best fit) and the dashed lines (dispersion) show the empirical relations between
various line ratios and the gas metallicity (see Fig.~\ref{fig_cal}).
The green errorbars show the observed
ratios (along the Y-axis) and the best-fit metallicity with uncertainty (along the
X-axis); the blue cross shows
the de-reddened ratios, by adopting the best-fit extinction; the red line shows the projection
of the 1$\sigma$ uncertainty of the fit in the top-left panel.}
  \label{fig_met_av_stack}
\end{figure*}

Since no single method is capable of providing a calibration of the
strong line diagnostics over the wide metallicity range required to sample the evolution
of galaxies through the cosmic epochs (7.7$<$12+log(O/H)$<$9.1), in this
paper we have to combine two different methods depending on the metallicity range.

At low metallicities ($\rm 12+log(O/H)<8.35$) we use calibrations of
the strong-line diagnostics based on the $\rm T_e$ method, which is
regarded as reliable at low metallicities and free of problems
related to temperature fluctuations and gradients
\citep{stasinska05,bresolin06b,bresolin07b}. We use the sample
of 259 low metallicities galaxies gathered by \cite{nagao06} for
which measurement of the auroral line [OIII]4636 is available.
The metallicity determination based on the $\rm T_e$ method is
detailed in \cite{nagao06}. The inferred metallicities for this sample
are in the range $\rm 7.1 < 12+log(O/H) < 8.3$. The inferred empirical relations
between metallicity and various strong-line diagnostics
are shown in Fig.~\ref{fig_cal} with blue squares. Note that
the $\rm T_e$ method is only used to calibrate the strong line diagnostics in
local galaxies (and at low metallicities), but it will not be used to directly
measure metallicities in our high-z objects, since the [OIII]4636 line
is too faint to be detected.

At $\rm 12+log(O/H) > 8.35$, where the $\rm T_e$ method is know to fail (as discussed
above), we have to rely on photoionization models. More specifically, we adopt the
calibrations provided by the models in \cite{kewley02}, aware of the caveats discussed above,
and in particular that these models are valid only at $\rm 12+log(O/H) > 8.4$.
We then used
the data of star forming galaxies in SDSS DR4, by adopting the same
constraints on the signal-to-noise discussed in \cite{nagao06}, and by
also considering only objects with $\rm log([NII]6564/[OII]3727])>1.2$,
as recommended by \cite{kewley02}. This selection results in a total of
22,482 objects. After determining the gas metallicity for each object
with the \cite{kewley02}
method, we derived the empirical relations with various strong-line diagnostics
as shown in Fig.~\ref{fig_cal} (black dots).

The relations obtained by combining both the low metallicity and the high
metallicity samples were fitted with a polynomial curve. To avoid
the fit to be dominated by the regions containing the largest number of
objects (i.e. by the SDSS sample at $\rm 8.7 <12+log(O/H)<9.1$), we 
divided the relations in metallicity bins (generally spaced by 0.1~dex), we
derived the median and dispersion of the strong-line flux ratio and of the
stellar masses within each bin, and then
fitted the polynomial function to these medians.
In most cases
a second order or a third order polynomial is appropriate to describe the
relation over the full metallicity interval. However, in some cases
a fourth order polynomial is required. The general functional form for describing
the strong-line metallicity calibration is therefore:
\begin{equation} \label{eq_met_cal}
\log{R} = c_0+c_1x+c_2x^2+c_3x^3+c_4x^4
\end{equation}
where $\log{R}$ is the logarithm of the strong-line ratio,
and $x$ is the metallicity relative to solar
($x=\log{(Z/Z_{\odot})}=12+\log{(O/H)}-8.69$, \cite{allende01}). The coefficients $c_0-c_4$ for
each strong-line ratio are listed in Tab.~\ref{tab_met_cal}, and the
resulting best fit polynomials are plotted with a solid red curve in
Fig.~\ref{fig_cal}.

\subsection{Determination of the gas metallicity}\label{sec_met_det}

In the redshift range 3$<$z$<$3.7 investigated in this paper [OII]3727 and (not always)
[NeIII]3870 are observable in the H band, while H$\beta$ and [OIII]5007 are observable in
the K band. This allows us to use five metallicity diagnostics (not all of them independent of
each other), namely: $\rm R_{23}=([OII]3727+[OIII]4959+[OIII]5007)/H\beta$,
$\rm [OIII]5007/H\beta$, $\rm [OII]3727/H\beta$, $\rm [OIII]5007/[OII]3727$,
$\rm [NeIII]3870/[OII]3727$. The relationship between these ratios and the gas metallicity,
along with their dispersion, are shown in Figs.~\ref{fig_cal} and \ref{fig_met_av_stack}.
Each of these diagnostics has advantages and disadvantages. For instance,
$\rm [OIII]5007/H\beta$ is essentially unaffected by dust reddening, but it has a double
metallicity solution for each value of this ratio. $\rm [OIII]5007/[OII]3727$ has a monotonic
dependence on metallicity, but it is potentially affected by dust reddening and it is also
affected by a larger dispersion.
$\rm [NeIII]3870/[OII]3727$ is both a monotonic function of metallicity and little affected
by dust reddening, nonetheless
[NeIII] is generally the faintest of all these lines and the most
difficult to detect. However, if these various diagnostics are used simultaneously, then
it is possible to both account for dust reddening and remove the ambiguity of double
solutions. Essentially, if all the diagnostics are used then only some
combinations of metallicity and dust reddening are allowed by the data.

More specifically, we selected the following independent metallicity diagnostics:
[OIII]5007/H$\beta$,
[OIII]5007/[OII]3727 and (when available) [NeIII]3870/[OII]3727 (note that the ratio
[OIII]5007/H$\beta$ is very similar to the ``classical'' R$_{23}$ parameter at the low
metallicities investigated by us, since at low metallicities [OIII]/[OII] is high).
We then determined the best pair of metallicity
and extinction that minimizes the $\chi ^2$ in the three corresponding diagrams,
both by including the measurement errors and the dispersion of each calibration diagram.
In practice, since [NeIII] is often undetected or not observable,
the metallicity is mostly determined
through the [OIII]5007/H$\beta$ ratio, while the [OIII]5007/[OII]3727 ratio is used
to discriminate which of the two metallicity solutions for the [OIII]5007/H$\beta$ ratio
applies and also to provide some constraints on the dust extinction (although the latter has
generally negligible effect on the metallicity determination since [OIII]5007/H$\beta$ is
insensitive to reddening). \footnote{ 
For the $\rm A_V$-metallicity fit we
used the Milky Way extinction curve \citep{cardelli89} ($\rm R_V=3.1$), which is often preferred
for the nebular lines.}

As an example, we show the results of this method in
Fig.~\ref{fig_met_av_stack} in the case of the composite spectrum.
The upper left panel shows the
best solution (blue cross) and the 1$\sigma$ confidence level (red curve,
obtained from solutions with $\Delta \chi ^2 = 1$)
in the $\rm A_V$--metallicity plane. In the other panels the black solid line
(best fit) and the dashed lines (dispersion) show the empirical relations between
various line ratios and the gas metallicity (Fig.~\ref{fig_cal});
the green errorbars show the observed
ratios (along the Y-axis) and the best-fit metallicity with uncertainty
(along the X-axis); the blue cross shows
the de-reddened ratios, by adopting the best-fit extinction; the red line shows the projection
of the 1$\sigma$ uncertainty of the fit obtained in the top-left panel.
It can be noted that the
extinction is subject to a large uncertainty, but the metallicity is
relatively well constrained.
The metallicities resulting from the procedure discussed above are listed for all
objects and for the composite spectrum in
Tab.~\ref{tab_met}. 

\subsection{Evolution of the metallicity diagnostics}\label{sec_exc}

One of the main worries when using strong emission line diagnostics in high-z sources
is that the empirical calibrations are obtained by using local sources. Since the
dependence of the strong lines ratios on metallicity involves also other dependences
(e.g. on the ionization parameter, on the shape of the ionizing radiation,
on the gas density),
the evolution of the average galaxy properties (e.g. SFR, compactness) with redshift may affect the
calibration of the metallicity diagnostics. It is very difficult to investigate this issue,
since in principle one would need to obtain an empirical calibration at high-z by observing
primary metallicity tracers (e.g. [OIII]4636), which are however extremely faint.
An alternative way is to construct
diagrams which are sensitive to the excitation mechanism, by disentangling the dependence
on metallicity, and verify whether the line excitation conditions change with redshift.

Such a test was performed, at lower redshifts, by
investigating the diagram [OIII]/H$\beta$ versus [NII]/H$\alpha$ for sources where
all of these lines are observable \citep{shapley05,erb06,liu08}. It is found that a fraction of sources
at z$\sim$1--2 are offset with respect to the sequence described by local HII galaxies,
and displaced towards the AGN locus. An interpretation is that at least some of these
sources are affected by some AGN contribution \citep{liu08},
which were not excluded from the sample because elusive in the UV rest-frame spectra.
Other sources may be characterized by truly different physical conditions with respect
to local HII regions, and in particular higher ionization parameter.
However, even in the latter cases both \cite{brinchmann08} and \cite{liu08} find that the
calibration of the strong line metallicity diagnostics do not deviate by a large amount
with respect to local HII galaxies. In particular, they find deviations by only about 0.1~dex
(or less) in terms of metallicity calibration, depending on the specific diagnostic adopted.

For what concerns the sources at z$\sim$3.5 in AMAZE,
the available emission lines allow us to construct the
so-called BPT diagram \citep{baldwin81}, i.e. [OIII]5007/H$\beta$
versus [OIII]5007/[OII]3727. As discussed in \cite{dopita06} this diagram
is strongly degenerate in terms of
metallicity, but it is sensitive to both the ionization parameter
and the hardness of the ionizing source.
Within the observational uncertainties,
the sources in our AMAZE sample do not deviate from the sequence of local HII galaxies on the
BPT diagram, suggesting that the excitation conditions do not differ significantly
from the local galaxies used to calibrate the strong metallicity diagnostics.

  \begin{figure*}[!ht]
   \centering
   \includegraphics[width=18.5cm]{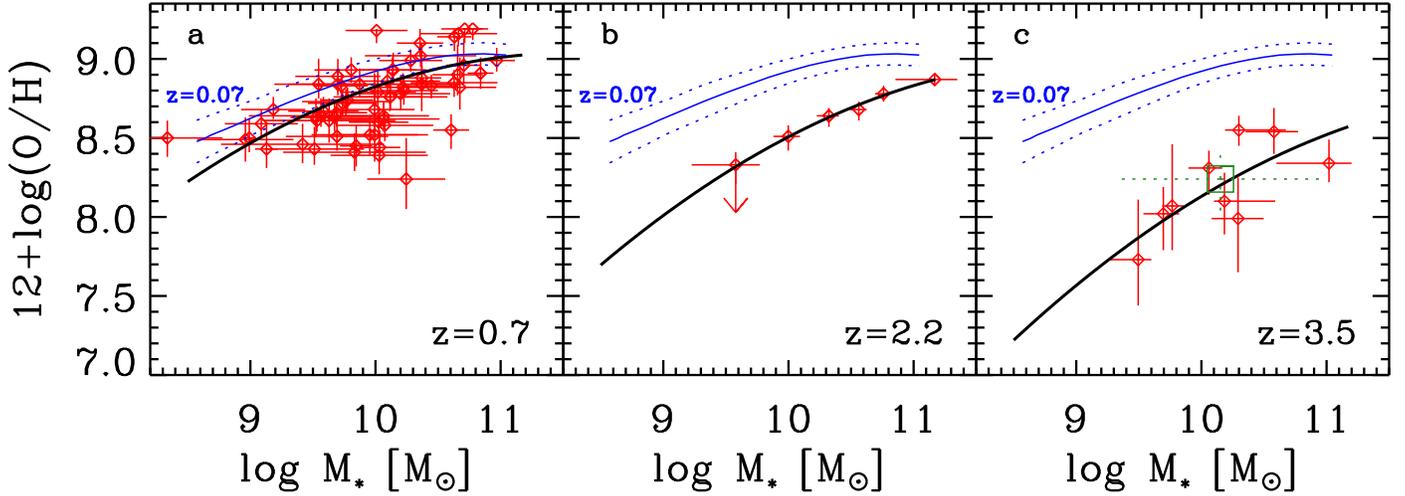}
\caption{Mass-metallicity relation observed at different redshifts.
The blue, solid and dotted lines indicate the
mass-metallicity relation and its dispersion observed at z$\sim$0.07, as
inferred by \cite{kewley08}.
The red diamonds with errorbars show the mass-metallicity relation at different
redshifts traced by individual objects (or by stacked spectra in the case
of panel {\it b}, z$\sim$2.2). In panel {\it c} (z$\sim$3.5) the green square with dashed
errorbars is the composite spectrum (which has been assigned the median mass of the sample,
but slightly offset to avoid a confusing overlap with an individual source).
Black, solid lines show the analytical function
Eq.~\ref{eq_mz_evol} with the best fitting parameters listed in Tab.~\ref{tab_mz_evol}.
}
  \label{fig_mz}
\end{figure*}

\section{Stellar masses}
\label{sec_mass}

To derive the stellar masses for the LBGs in the AMAZE sample,
we used an approach based on broad-band spectral fitting technique.
Broad-band photometric data for the sources in the CDFS were collected
from the GOODS--MUSIC multiwavelength catalog \citep{grazian06}. This
catalog provides photometric data in 14 spectral bands (from UV to
the Spitzer-IRAC bands), and it has been recently updated to include
the Spitzer--MIPS data at 24$\mu$m. For the LBGs in 
\cite{steidel03}, optical photometric data (U,G,R,I) were extracted
from the publicly available images \citep{steidel03}, while
Spitzer IRAC and MIPS data were obtained from the Spitzer archive;
the photometry extraction was performed following the same methods
described in \cite{grazian06}.

The SED fitting technique adopted here is the same as in previous
papers \citep{fontana06,grazian06,grazian07}, and
similar to those adopted by other groups in the literature
\citep[e.g.][]{dickinson03,drory04,pozzetti07}. This technique is based on
comparing the observed multicolor distribution of each object and a
set of templates, computed with standard spectral synthesis models
(see below) and chosen to
broadly encompass the variety of star--formation histories, ages,
metallicities, and extinction of real galaxies. More specifically,
we considered exponentially decaying SFR with e-folding times ranging
from 0.1 to 15 Gyr. We used the Salpeter
IMF ($M_{min}=0.1M_{\odot}$ and $M_{max}=65M_{\odot}$), ranging over a
set of metallicities (from $Z=0.02 Z_\odot$ to $Z=2.5 Z_\odot$) and
dust extinction ($0<E(B-V)<1.1$, with a \cite{calzetti00}
attenuation curve, which
is generally more appropriate for the stellar component).
For each model of this grid, we computed the expected magnitudes in
our filters set and found the best--fitting template with a standard
$\chi^2$ normalization. The stellar mass and other best--fit
parameters of the galaxy, like SFR, age, and dust extinction, are
fitted simultaneously to the actual SED of the observed galaxy.
The metallicity of each galaxy is
fixed to value closest to the one determined by us through the
nebular lines (Tab.~\ref{tab_met}).

The stellar mass derived here is subject to uncertainties and biases
related to the synthetic libraries used to carry out the fitting of
the galaxy SEDs.  In general, the stellar mass turns out to be
the least sensitive parameter to variations of the input model assumptions, and the
extension of the SEDs to mid-IR wavelengths (near-IR rest-frame)
with IRAC greatly reduces
the formal uncertainties on the derived stellar masses, as shown in
\cite{fontana06}.
The uncertainties in the stellar mass are derived as follows: we
compute the 90\% confidence level on the mass by scanning the $\chi^2$
levels, fixing the redshift and the metallicity for each galaxy but
allowing the other parameters (SFR, age, dust extinction) to change.

Age and star formation rate are more uncertain
parameters to derive. In some cases we formally obtain best-fit ages below $\sim$50~Myr, which
are below the dynamical timescales for the star forming regions in these systems
\citep{shapley01}. Moreover, the conversion between UV luminosity and SFR becomes highly
non-linear below this age. As a consequence, we decided to restrict the allowed
ages to $>$50~Myr. However, this choice may only affect the inferred 
SFR, while the determination of the stellar mass is essentially unaffected,
as discussed above.

For what concerns the library of spectral synthesis models we adopt both
those provided by \cite{bruzual03} (hereafter BC03) and those by \cite{maraston05} (hereafter M05).
The resulting
stellar masses are tabulated for both cases in Tab.~\ref{tab_phys}.
The masses inferred by using the M05 models are preferred, since they
take into account the contribution by TP-AGB stars, and may differ from the stellar masses
obtained with BC03 by even a factor of two, especially in older stellar systems.
However, previous works on the
mass-metallicity relation at lower redshift adopted the BC03 templates. Therefore,
when comparing our results with the previous works at lower redshifts
we will adopt for consistency the masses obtained with the BC03 templates.
In Tab.~\ref{tab_phys} we also list the SFR, age and reddening inferred by the
SED fitting (adopting templates by BC03, again for a consistent comparison with previous
works).

\section{The mass--metallicity relation at high redshift} \label{sec_mz}

\subsection{Comparing the mass-metallicity relation at different redshifts} \label{sec_mz_z}

Different studies of the mass-metallicity relation at various redshifts have employed
different diagnostic lines and different calibrations. As discussed in \S\ref{sec_met} and
more extensively in \cite{kewley08}, the mismatch between the different calibration scales may
introduce artificial evolutionary effects of the mass-metallicity relation.
Therefore, it is important that
different strong-line diagnostics used in different surveys are cross-calibrated in a consistent way.
The relations obtained in \S\ref{sec_met} provide such a common
cross-calibration between different strong-line diagnostics on the same metallicity scale.
In this section we apply (when required) the correction to the metallicities inferred by past
surveys at lower redshift to match our metallicity scale.
We also apply corrections to the mass scale to account for the different IMF's adopted
by previous works.

As discussed in \S\ref{sec_intro} the local (z$\sim$0.1) mass-metallicity relation
was derived by \cite{tremonti04} by using SDSS spectra from DR2. \cite{kewley08} re-determined
the local mass-metallicity relation by using SDSS spectra from DR4 by setting tighter limits on the
redshift range (0.04$<$z$<$0.1) so that the projected SDSS fiber covering factor is
$>$20\%  of the total photometric g$'$-band light, and also to minimize
incompleteness effects at higher redshifts.
The resulting median redshift of their sample is $\sim$0.07.
\cite{kewley08} calibrate the metallicities with the \cite{kewley02} method, which is the same
adopted by us at 12+log(O/H)$>$8.35, hence no additional correction is required to match
our metallicity scale. The only correction to apply
is for the stellar masses, since \cite{tremonti04} and \cite{kewley08}
adopt a different IMF \citep{kroupa01}.
We calculate that the masses in \cite{kewley08} must be multiplied by a factor of 1.17 to comply
with our IMF (note that the IMF's differs not only in terms of shape but also
in terms integration limits).
The thin blue solid lines in Fig.~\ref{fig_mz} show the \cite{kewley08} mass-metallicity relation
corrected as discussed above. The blue, dotted lines indicates the 1$\sigma$ dispersion of the same
relation.

At 0.4$<$z$<$1 we use the results by \cite{savaglio05}. For consistency with our calibration scale
we re-determine the metallicities for each object in their sample by applying the same procedure
described
in \S\ref{sec_met_det} to the line fluxes tabulated by them. We exclude from their sample objects
without K-band data, since in these cases the stellar mass uncertainties are too large.
We also have corrected the stellar masses in \cite{savaglio05} by a factor of 1.4 to comply
with the IMF adopted by us.
The resulting mass-metallicity relation at z$\sim$0.7 is shown with red diamonds and errorbars in
Fig.~\ref{fig_mz}a.

At z$\sim$2.2 we use the results by \cite{erb06}, who infer the metallicity of LBG's at this
redshift through the [NII]/H$\alpha$ ratio measured in {\it stacked} spectra. Also in this case
we re-determine the metallicity in each mass bin by using the [NII]/H$\alpha$ metallicity
calibration obtained in \S\ref{sec_met}, to be consistent with the calibrations adopted by us.
The stellar masses in \cite{erb06} have to be corrected by a factor of 1.4 to comply
with the IMF adopted by us.
The resulting mass metallicity relation at z$\sim$2.2 is shown with red diamonds and errorbars in
Fig.~\ref{fig_mz}b.

Finally, Fig.~\ref{fig_mz}c shows the mass-metallicity relation inferred from the initial sample of
nine AMAZE sources
at z$\sim$3.5. The red diamonds with solid errorbars are individual objects. 
The green square with dashed errorbars is
the composite spectrum. In this plot, for consistency with the other
works at lower redshifts, we use the stellar masses inferred with the BC03 templates.

Previous versions of our mass-metallicity relation z$\sim$3, presented in our
previous preliminary works \citep{maiolino07a,maiolino07b},
were slightly different because of lower S/N spectra and
also because we used different calibrations (both for metallicity and stellar masses).

For a more straightforward comparison of the mass-metallicity relation at different redshifts,
it is useful to describe these relations by fitting them with the same functional form.
In order to minimize the number of free parameters we find statistically satisfactory
an approach similar
to \cite{savaglio05}: the quadratic function fitting the local mass-metallicity relation
is shifted in mass and in metallicity
to provide the best fit of the mass-metallicity relation at various redshifts.
More specifically we adopt the following description of the mass-metallicity relation:
\begin{equation}\label{eq_mz_evol}
\rm 12+log(O/H) = -0.0864~(log M_* - log M_0)^2 + K_0
\end{equation}
where $\rm log M_0$ and $\rm K_0$ are determined at each redshift to obtain the 
best fit to the observed data points\footnote{Note that, to avoid the fit to be dominated
by the mass intervals with largest number of objects, we first obtained metallicity medians
within mass bins and then fitted the resulting values with Eq.~\ref{eq_mz_evol}.},
as listed in Tab.~\ref{tab_mz_evol}.

\subsection{Aperture effects}
\label{sec_apert}

Since galaxies, and especially disk galaxies, are often characterized by metallicity gradients
(the metallicity decreasing towards the outer regions),
a possible caveat when comparing metallicities at different redshifts 
is the different aperture projected on the source. In particular, at high redshift spectroscopic
observations are likely to sample most of the galaxy, while at low redshift and in local galaxies
the spectroscopic aperture samples mostly the central higher metallicity region. This effect
may mimic a metallicity evolution. 

When comparing surveys at high redshifts this should be a minor issue, since the projected
apertures on the sources are not very different.
At z$\sim$0.7, the adopted aperture of 0.75-1.3$''$ in \cite{savaglio05}
corresponds to about 5.6-9.3~kpc; at z$\sim$2.2 \cite{erb06} adopt an aperture of 0.76$''$
corresponding to about 7.2~kpc. Our aperture of 0.75$''$ at
z$\sim$3.5 corresponds to about 6~kpc.

However, aperture effects may be more serious for the local sample.
At the median redshift of 0.07 the SDSS fiber size (3$''$) has a median size of 4~kpc,
and a median covering factor $\sim$34\% relative to the total g'-band light, in contrast
with a covering factor of $\sim$70\% at z$\sim$3.5. The aperture effect is stronger for local
high mass galaxies, which are generally bigger and for which the covering factor reach values
as low as 20\%. The problem of a differential aperture effect as a function of galaxy mass
may also affect the shape of the local mass-metallicity relation, making the observed relation
steeper than it actually is \citep{kewley08}. However, the absolute magnitude of this effect
in the local sample is estimated to be at most $\sim$0.1--0.15~dex \citep{kewley08}, which is
significantly lower than the metallicity evolution observed at high redshift, at least at z$\sim$3.5.

Additional issues related to aperture effects will be discussed in the context of the comparison
with models in \S\ref{sec_mod}.

\subsection{Selection effects}\label{sec_bias}

When comparing the mass-metallicity relation at z$\sim$3.5
inferred from LBG's with that inferred from local (or lower redshift) samples of star forming
galaxies, one must be aware
that we are comparing different classes of objects, which are not necessarily
linked from an evolutionary point of view. As a consequence, the evolution of the mass-metallicity
relation inferred in this paper should be regarded as the evolution of the
mass-metallicity relation of
galaxies representative of (or contributing significantly to) the density of star formation at
each epoch, and {\it not} the evolutionary pattern of individual galaxies. This issue will be
further discussed in the next sections.
In this section we mostly investigate whether 
galaxies in our sample are representative of star forming galaxies at z$\sim$3.5.

Stellar masses of LBGs at z$\sim$3 were measured
by \cite{shapley01}\footnote{Note that \cite{shapley01}
did not have access to Spitzer-IRAC data, and therefore their masses may be subject to significant
uncertainties.}, but for a subsample of galaxies
about one magnitude brighter (in R-band) than the parent sample of \cite{steidel03}.
\cite{shapley01} obtain a median stellar mass of $\rm 2.4\times 10^{10}~M_{\odot}$.
The galaxies in our sample (Tab.~\ref{tab_phys} and \S\ref{sec_mass}) have a median stellar mass of
$\rm \sim 1.6\times 10^{10}~M_{\odot}$, which is close to the value obtained by \cite{shapley01}
for their large LBG sample.
In any case, the distribution of stellar masses is not a concern, since the 
mass is one of the two variables
that we are mapping on the mass-metallicity relation: even if we had a bias in terms of stellar mass,
this would simply imply that we preferentially populate the diagram in a certain mass range, making
the estimation of the mass-metallicity relation more uncertain in other mass ranges (because
under-populated), but not biased.

Biases in terms of star formation rate are of a greater concern. Our selection requirement that sources
must have a highly reliable spectroscopic redshift may bias our sample towards sources with strong UV
continuum or strong Ly$\alpha$, hence higher than average SFR. Tab.~\ref{tab_phys} shows the SFR
inferred from the rest-frame UV continuum of the sources in our sample (see \S\ref{sec_mass} for
details), from which we infer a median SFR of $\rm \sim 100~M_{\odot}~yr^{-1}$. This is similar
to the median SFR ($\rm 90~M_{\odot}~yr^{-1}$) of LBGs at z$\sim$3 obtained by
\cite{shapley01} (who adopted a similar approach as ours to estimate the SFR from the UV continuum).
However, since the latter work is biased towards slightly brighter optical
magnitudes, the median SFR of the LBGs at z$\sim$3 in
the parent sample of \cite{steidel03} is probably somewhat lower.
Since the 
median stellar masses are similar, then the specific star formation rate (SSFR) in our sample
may be somewhat higher (up to a factor of 2) than in the LBG sample of \cite{steidel03}.
\cite{ellison08} investigated the effect of the SSFR on the metallicity in
local galaxies. They
found that, for a variation of the SSFR by a factor of two,
the metallicity varies by less than $\sim$0.1 dex
in low mass galaxies ($\rm M_{*}<10^{10}M_{\odot}$), while no metallicity variations are found
in massive galaxies ($\rm M_{*}>10^{10}M_{\odot}$). As a consequence, a possible bias of
our sample in terms of SSFR relative to the LBG sample of \cite{steidel03} should not affect
significantly the inferred mass-metallicity relation.

Another possible source of bias is that LBGs are selected through their UV rest-frame colors,
hence they miss any population of heavily reddened star forming galaxies. \cite{reddy07}
estimate that, due to this selection effect, LBG's represent $\sim 47$\% of the population
of star forming galaxies at z$\sim$3 with $\rm R<25.5$
\citep[see also ][]{hopkins06}. Dust reddened galaxies
are naively expected to be more metal rich
\citep[since metallicity and dust content correlate, ][]{hunt05}. On the contrary,
\cite{rupke08} have shown that, at least at low-z,
dusty galaxies (IR-selected) are characterized by gas metallicities lower than optical- and
UV-selected galaxies (probably due to metal poor gas infalling from the outskirts in
merging/interacting systems). Recently \cite{caputi08} have found a similar effect 
in dusty galaxies at intermediate redshift (i.e. metallicities lower than in optically
selected galaxies). High-z dusty objects may behave
differently. However, if the same phenomenon is also present at high-z, then 
LBG's would provide an upper limit to the metallicity of galaxies at z$\sim$3. However,
we do not speculate further on the properties of star forming galaxies that are not sampled by LBGs,
and we simply emphasize that the results presented in this paper
only apply to about half of the star forming galaxies at z$\sim$3, i.e. those
UV-selected.

\begin{figure}[!t]
   \centering
   \includegraphics[width=8cm]{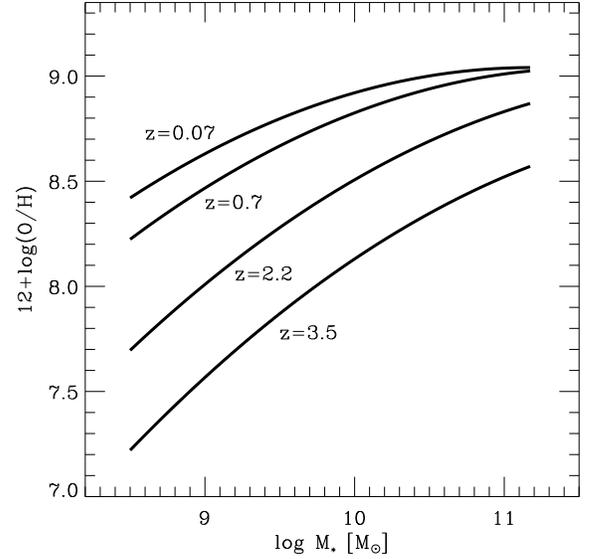}
\caption{Comparison of the mass-metallicity relation observed at different redshifts,
as parametrized by the analytical function Eq.~\ref{eq_mz_evol} and coefficients
in Tab.~\ref{tab_mz_evol}.
}
  \label{fig_mz_z}
\end{figure}

  \begin{figure}[!ht]
   \centering
   \includegraphics[width=8.5cm]{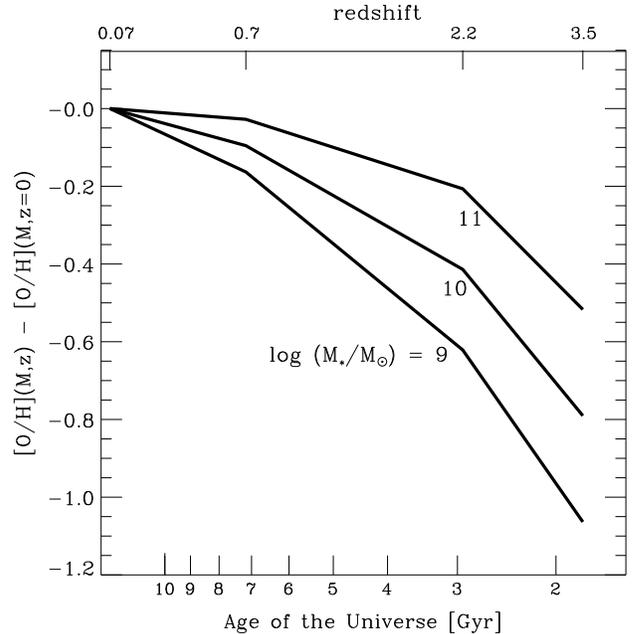}
\caption{Average metallicity of star forming galaxies as a function of the cosmic age
of the Universe, relative to local galaxies,
for three different families of galaxies with different stellar masses
($\rm M_*/M_{\odot} = 10^9,~10^{10},~10^{11}$). Ages are sampled at the redshifts where
observations are directly available. The average metallicity for each stellar mass was
inferred by using the parametrization given by Eq.~\ref{eq_mz_evol} and Tab.~\ref{tab_mz_evol}.
}
  \label{fig_mz_evol}

\end{figure}

  \begin{figure*}[!t]
   \centering
   \includegraphics[width=12cm]{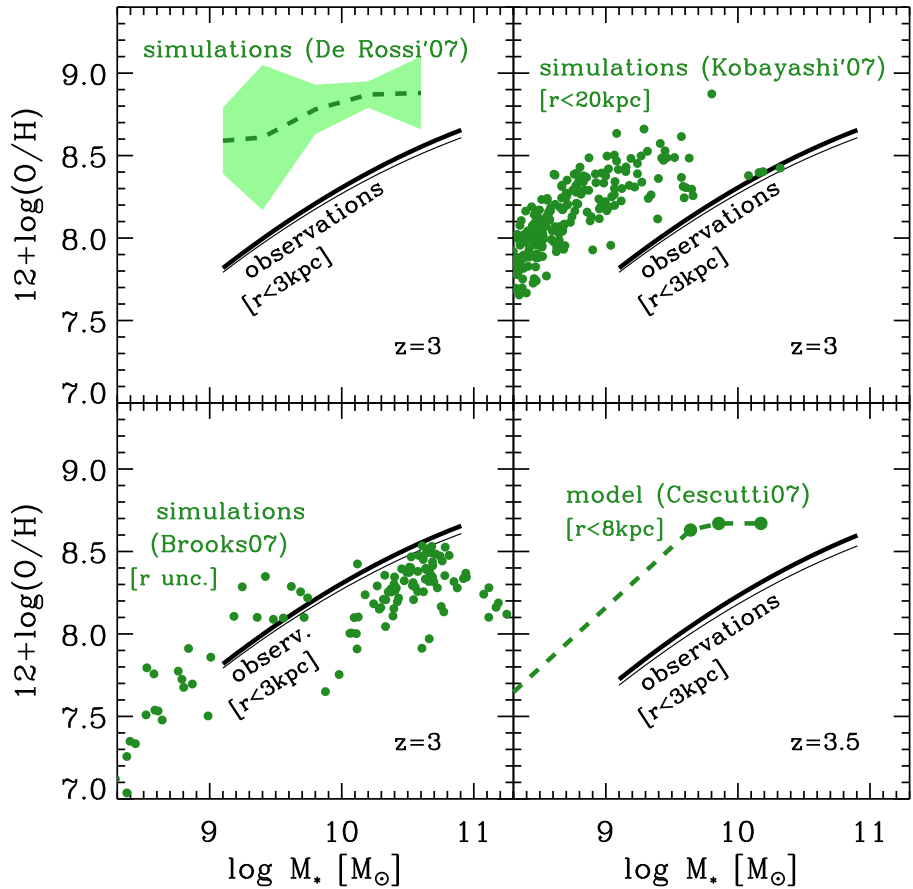}
\caption{Comparison between models/simulations predictions for the mass-metallicity relation at
z$\sim$3 and the results of our observations. The predictions obtained by models
and simulations are shown with green points and dashed lines (the
aperture used in the models/simulations is reported when available).
The solid, black lines show the mass-metallicity relation observed at the same
epochs (interpolated to the same, exact redshifts of the simulation by using Eq.~\ref{eq_mz_evol}
and Tab.~\ref{tab_mz_evol}).
The thin black line shows the observed mass-metallicity relation by using
the masses estimated with the BC03 templates, while the thick line is the inferred relation when using
the M05 templates.
}
  \label{fig_models}
\end{figure*}

\subsection{The evolution of the mass-metallicity relation}

The evolution of the mass-metallicity relation is summarized in Fig.~\ref{fig_mz_z}, where we
plot the best fits resulting from Eq.~\ref{eq_mz_evol} both at z$\sim$3.5 from the AMAZE
survey, and at lower redshifts from previous surveys.

Figs.~\ref{fig_mz}--\ref{fig_mz_z} highlights a clear evolution of the
mass-metallicity relation of star forming galaxies
through the cosmic epochs. As already discussed in \S\ref{sec_bias}, this evolution
should {\it not} be seen as the evolutionary sequence of individual objects, since at each
redshift the various surveys are sampling different classes of star forming galaxies,
which are not necessarily each other progenitors.
The trend observed in Figs.~\ref{fig_mz}--\ref{fig_mz_z}
should be regarded as the evolution of the mass-metallicity relation of galaxies
dominating (or contributing significantly) the star formation density at each epoch.

At $\rm M_*\sim 10^{10} M_{\odot}$
the metallicity at z$\sim$2.2 is lower by a factor of about 2.5 with respect to local
galaxies. Even if highly significant, such metallicity
decrease is modest if one considers that from z$=$0 to z=2.2 the elapsed time is
$\sim$11~Gyr, i.e. about 75\% of the age
of the universe. From z$\sim$2.2 to z$\sim$3.5 the average metallicity of galaxies decreases by
another factor of about 2.5. However, the latter evolution is much stronger, and faster.
Indeed, such a metallicity variation occurs on a much shorter time scale, only $\sim$1~Gyr.
This effect is shown in more clearly Fig.~\ref{fig_mz_evol}, where the the average metallicity
of star forming galaxies is plotted
as a function of the age of the universe, for different classes of galaxies with
different stellar masses, by exploiting the analytical function in Eq.~\ref{eq_mz_evol}
(note that, as discussed above, this figure does not provide the evolution of individual galaxies).
We further note that the evolution is strong even in massive galaxies.
Clearly, z$\sim$3.5 is an epoch of major action
for the evolution of galaxies, both in terms of star formation and chemical enrichment,
even for massive systems.

We note that a similar strong evolution of the metallicities at z$>$3 was obtained by
\cite{mehlert06} by investigating the {\it stellar} metallicities of a few galaxies with
bright UV continuum. Their result was however affected by uncertainties on the absolute
calibration of the stellar metallicity tracers.
It is also interesting to note that the metallicities obtained by us at z$\sim$3.5 
are in fair agreement with those expected at the same redshift by \cite{panter08}, who
inferred the metallicity evolution of galaxies by 
modelling its ``fossil'' spectral signatures in local galaxies.

The additional interesting result is the indication of a differential, mass-dependent
evolution of the metallicity. In particular, the metallicity evolution in low mass
systems appears stronger than in massive galaxies (Figs.~\ref{fig_mz}--\ref{fig_mz_evol}).
This finding requires more
statistics to be confirmed at z$\sim$3.5 (to come with the completion of the AMAZE project).
However, the evolution of the slope of the mass-metallicity relation (relative to the local slope)
is significant even at z$\sim$0.7 \citep[as already noted by ][]{savaglio05,kobulnicky03}
and at z$\sim$2.2. A detailed investigation
of differential selection effects as a function of stellar mass is required to rule out
that observational biases are not affecting the slope of the mass-metallicity relation at each epoch.
However, if confirmed, such mass-dependent evolution of the metallicity
can be regarded as the ``chemical'' version of the galaxy downsizing:
high mass galaxies reach high metallicities at high redshift, on short timescales, while
low mass systems enrich their ISM over a prolonged period of time, extending to the current epoch.

\subsection{Comparison with models of galaxy evolution}
\label{sec_mod}

As mentioned in the introduction there is an intense theoretical activity aimed at interpreting
the nature and origin of the mass-metallicity relation, and also at providing predictions on the
expected mass-metallicity relation at high redshift.
Comparing the models predictions with the observational results in a consistent way
is not simple. Indeed, theoretical models predict a variety of galaxy populations, spanning
a wide range of properties (e.g. in terms of SFR), while
observations are limited to samples matching the survey selection criteria.
Moreover, generally theoretical works provide the metallicity integrated over large apertures,
including most of the galaxy, while observational metallicity measurements are generally obtained
within a smaller aperture. In the future a collaborative effort with various theoretical groups is
planned to match the outcome of models and simulations to the observational
selection effects.
However, a preliminary comparison with the already published theoretical results is instructive
to infer some initial constraints on galaxy evolutionary models.

\cite{savaglio05} interpreted the evolution of the mass-metallicity relation from z$\sim$2.2
to z$=$0 through a closed-box model with an exponentially decaying SFR $\propto e^{-t/\tau}$,
where the $\it e$-folding time $\tau$ decreases as a function of galaxy mass. However the inclusion
of our data at z$\sim$3.5 makes this model not suitable. It is difficult
to simultaneously fit the observed mass-metallicity relations at z=0.1, z=0.7, z=2.2 and z=3.5
with a simple closed-box model, unless more complex scenarios of the SF history are envisaged.
Moreover, an exponentially decaying SFR, with the {\it e}-folding times provided by
\cite{savaglio05} makes the SFR extrapolated to local massive systems well below
1~$\rm M_{\odot}~yr^{-1}$, i.e. these should be local quiescent galaxies (probably massive
elliptical), which cannot
be representative of the local star forming galaxies used to derive the local mass-metallicity
relation in \cite{tremonti04} and \cite{kewley08}. Finally, it is unlikely that the closed-box
scenario applies to LBGs, which are characterized by strong, unbound winds
\citep[e.g.][]{pettini02}.

Within the framework of the hierarchical
models of galaxy evolution, \cite{derossi07} performed numerical hydrodynamical simulations
enabling them to provide detailed predictions on the evolution
of the mass-metallicity relation at various epochs.
In the top-left panel of Fig.~\ref{fig_models} the dashed green line shows the mass-metallicity
relation predicted at z$=$3
by the simulations of \cite{derossi07} (therein Tab.~3),
while the shaded areas give the dispersion inferred from the same simulations \citep[note that ][ use
the same IMF adopted by us, therefore no further correction is required]{derossi07}.
Since we do not have the observational data at exactly the same epoch of the simulations (z$=$3),
we interpolate the observed mass--metallicity to
the same epoch of the simulations by using Eq.~\ref{eq_mz_evol} and Tab.~\ref{tab_mz_evol}.
The {\it observed} mass-metallicity relation at z$=$3
is shown in Fig.~\ref{fig_models} with a thin, black solid line. 
We also show the mass-metallicity relation inferred by adopting
the stellar masses measured with the M05 templates (thick solid line).
Fig.~\ref{fig_models} shows a significant discrepancy between simulations
and observations. The discrepancy was also present at z$=$2 (although at a lower level),
as noted by \cite{derossi07}. They suggest that the inconsistency between simulations
and observations is due to the lack of significant SN feedback in their simulations, which would
remove metal enriched gas and lower the global gas metallicity.

However, the discrepancy with observations is present also for simulations that include the effect
of SN feedback. This is the case of the hierarchical, three-dimensional chemodynamical simulations
presented in \cite{kobayashi07}, which include the feedback from SNII and hypernovae.
They provide predictions of the mass-metallicity relation as a function of redshift.
The green points in the top-right panel of Fig.~\ref{fig_models} show the results
of such simulations at z=3 \citep[also ][ use
the same IMF adopted by us, therefore no further correction is required]{kobayashi07}.
Even in this case there is a discrepancy between simulations and observations.
The discrepancy is much reduced with respect to the \cite{derossi07} simulations, but still
significant at low masses ($\rm log(M_*/M_{\odot})<10$). At high
masses ($\rm log(M_*/M_{\odot})\sim 10.5$) simulations and observations are nearly consistent. However,
one should take into account that \cite{kobayashi07} extract the metallicities in the simulated
galaxies within a radius of $\rm r<20 kpc$, i.e. by including metal poor external regions,
while observations provide metallicities within a radius of $\rm r<3 kpc$. Aperture
effects are probably more important in large, massive galaxies. If the extraction
radius of the simulations is matched to the observations then the discrepancy probably increases
strongly also in the high mass region. 

The main problem of the latter simulations seems to be that the bulk of the chemical enrichment
occurs in small galaxies, yielding a steep metallicity evolution at masses
below $\rm <10^9~M_{\odot}$ on
the mass-metallicity plane. Then such evolved sub-units merge to form massive systems,
with little additional star formation (close-to-dry merging), implying little additional
enrichment (i.e. flatter evolution towards high masses). The discrepancy with
the observations
seems to imply that the evolution of galaxies at high redshift occurs through
the assembly of little evolved small galaxies, so that high-z 
objects with $\rm log(M_*/M_{\odot})> 9$
can exist with relatively low metallicities. In these systems most of the star formation
and chemical evolution occurs once they are already assembled into bigger systems.

The latter scenario is well described by the simulations presented in
\cite{governato07}, which model the evolution of disks within a
hierarchical framework. 
In these simulations a strong feedback,
due to SNe and to gas heating by the UV radiation,
prevents small galaxies to evolve significantly before merging into a bigger galaxy,
while the bulk of the chemical evolution and star formation occurs in the gas which has already
settled into the proto-disk.
\cite{brooks07} inferred the evolution of the mass-metallicity relation from these
simulations, whose prediction at z$=$3 is shown in the bottom-left panel of 
Fig.~\ref{fig_models} (Brooks, priv. comm.). The agreement with the observations
is good, although at high masses the simulations
tend to under-reproduce the observed metallicity. However the metallicities provided by
\cite{brooks07} are obtained without any constraints on the aperture (i.e. all cold gas);
if the information from the simulated galaxies is
extracted within our observational aperture ($\rm r<3~kpc$) then metallicities are
expected to increase (especially in large massive galaxies) and better reproduce the observations.
Yet, a potential problem of the \cite{governato07} and \cite{brooks07} models is that
their forming disks are characterized by modest star formation rates,
never exceeding $\rm \sim 20~M_{\odot}~yr^{-1}$,
while LBGs (except for a minority of them) are
characterized by significantly larger SFR, suggesting that they are in the process of rapidly
forming spheroids.

Detailed predictions on the mass-metallicity relation were also obtained by \cite{finlator08}
who used three-dimensional hierarchical simulations along with detailed outflows models.
They show that the evolution of the mass-metallicity relation out to z$\sim$2 can
be well reproduced if a ``momentum-driven wind'' model is incorporated. The predictions
of their model at z$\sim$3 reproduce reasonably well also the metallicity in massive galaxies
($\rm log(M_*/M_{\odot})\sim 10.5$) observed by us. 
However, their model predicts an
up-turn of the mass-metallicity slope at z$\sim$3 which is not observed by us.
Yet, the slope of the mass--metallicity relation is still poorly determined in our
data, due to the shortage of low-mass galaxies; we should wait for the completion of the
AMAZE program before claiming any significant inconsistency with the model in terms of slope
of the relation.

In the bottom-right diagram of Fig.~\ref{fig_models} 
we also show the mass-metallicity relation expected by the
double-infall models for the formation of galactic disks and dwarfs presented in \cite{chiappini01}
and \cite{cescutti07}. The green points show the mass-metallicity relation of spirals
according to these models, by tracing back their evolution until z$=$3.5.
The figure shows a significant discrepancy
between the model and the observations.
Such a discrepancy is not surprising, since LBG's at z$>$2 are probably
spheroids in the process of rapidly forming stars, and not spirals. More specifically,
in the double-infall model for disks the star formation
rate never exceeds a few times $\rm 10~M_{\odot}~yr^{-1}$, while the median SFR of LBGs
is $\rm 90~M_{\odot}~yr^{-1}$.

Summarizing, there are currently no models or simulations
that can satisfactory explain the mass-metallicity relation observed at z$\sim$3.
The closest match is probably with the simulations of \cite{governato07} and \cite{brooks07},
although even in these cases there are some discrepancies in terms of SFR.
The location of LBG's at z$\sim$3 on the mass-metallicity plane, along with
comparison with these models, suggest that
z$\sim$3 galaxies have been assembled through low mass systems
whose star formation efficiency was suppressed, hence which were little evolved.
The bulk of the star formation and of the chemical enrichment occurred
once small galaxies were already assembled
into bigger systems. In other words, most of the merging occurred before
most of the star formation.

  \begin{figure}[!t]
   \centering
   \includegraphics[width=8.5cm]{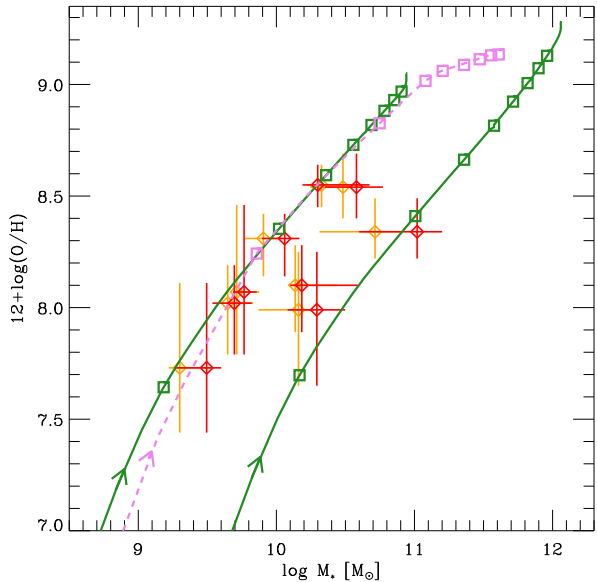}
\caption{Evolutionary tracks of {\it individual} galaxies on the
mass-metallicity diagram according to some
models for the formation of spheroids presented in \cite{granato04} (green solid lines)
and in \cite{pipino06} (violet dashed lines),
compared with the mass-metallicity relation observed
at z$\sim$3.5 (diamonds with errorbars). Squares mark the models temporal evolution
from 0.02 Gyr to 0.3 Gyr, in steps of 0.04 Gyr. Red symbols show the masses of the observed objects
inferred with the BC03
templates, while orange symbols are the masses inferred with the M05 templates.
}
  \label{fig_mono_ell}
\end{figure}

We conclude this section by comparing in Fig.~\ref{fig_mono_ell}
the location of z$\sim$3.5 galaxies on the mass-metallicity diagram with the evolutionary
tracks (as a function of time) expected for individual spheroidal galaxies,
according to the models
in \cite{granato04} and \cite{pipino06}.
These models
prescribe a nearly monolithic formation of elliptical
galaxies, where pristine gas collapses from the halo. In these models star formation is gradually
quenched as the galaxy evolves due to the feedback introduced by star formation and/or AGN activity.
The mass-metallicity tracks of these models are shown in Fig.~\ref{fig_mono_ell}
for different final stellar masses:
green solid lines and violet dashed line are for the \cite{granato04} model
and for the \cite{pipino06} model, respectively. The
observational data from the Amaze project are shown with diamonds and errorbars.
Red points are for masses inferred by using the
BC03 templates, orange points are for masses inferred by using the M05
templates.
The models can easily embrace the observed data points.
This comparison is not aimed at explaining the mass-metallicity
relation at z$\sim$3.5, since models of individual galaxies at different masses
must be convolved with the evolution of cosmic structures to
obtain a prediction of the mass-metallicity relation at any epoch.
Nonetheless, 
Fig.~\ref{fig_mono_ell} shows that the combination of mass and metallicities observed
in individual star forming galaxies at z$\sim$3.5 does not lie in a region difficult to populate
by models of individual galaxies;
on the contrary, individual observations can be easily explained in terms
of rapidly evolving, massive systems, through these simple models.
However, convolving these models
with the hierarchical growth of dark matter structures is required to verify whether they
can really explain the mass-metallicity relation at z$\sim$3.5.

\section{Summary and conclusions}

We have presented initial results of the AMAZE project, an ESO large programme
aimed at determining the mass-metallicity relation of star forming galaxies at z$>$3.
Near-IR spectra are being obtained with SINFONI, the VLT near-IR integral field spectrometer,
for a sample of Lyman Break Galaxies at 3$<$z$<$5.

Gas metallicities are inferred by using a combination of diagnostics involving nebular
lines observable in the H and K bands. To have a metallicity scale consistent with the results
obtained by previous surveys at lower redshifts, we derived new accurate calibrations
of various strong-line metallicity diagnostics spanning the wide range $\rm 7.1<12+log(O/H)<9.1$.

AGNs (which would affect and make unusable the metallicity diagnostics)
were carefully removed through a multiwavelength approach using X-ray, optical and mid-IR
data. In particular, the use of mid-IR (Spitzer-MIPS) data
allows us to discard even heavily obscured, Compton thick AGNs.

Stellar masses are inferred by fitting multi-band
photometric data with galaxy templates. Within this context crucial is the use of Spitzer-IRAC
data, which sample the rest-frame near-IR stellar light at 3$<$z$<$5.

In this paper we have presented results from an initial sample of 9 LBGs at 3.1$<$z$<$3.7.
Emission lines required to constrain the gas metallicity are detected in all sources.
From the mass-metallicity relation at z$\sim$3.5 inferred from this initial sample we obtain
the following results:

\begin{enumerate}

\item When compared with the results of
previous surveys at lower redshift (recalibrated to match our metallicity and mass
calibrations),
we obtain evidence for a clear evolution of the mass-metallicity relation as a function of redshift.
Since at different redshifts we are sampling different populations of galaxies, the observed
evolution should {\it not} be considered the evolutionary pattern of individual galaxies, but
as the evolution of the mass-metallicity relation of the dominant population of star forming
galaxies at each epoch.
The evolution of the mass-metallicity relation is faster at 2.2$<$z$<$3.5 than at later
epochs (z$<$2.2). This result indicates that z$\sim$3 is an epoch of major action for
the enrichment of galaxies, even in high mass systems

\item There are indications that the metallicity evolution is not constant with mass: low mass galaxies
evolve more strongly than massive systems. If confirmed, this result can be considered as
the chemical version of the galaxy downsizing, i.e. a scenario
where high mass galaxies reach high metallicities at high redshift, on short timescales, while
low mass systems enrich their ISM over a prolonged period of time, extending to the current epoch.
However, this finding is subject to uncertainties due to low statistics at low masses, and
must await additional data to be confirmed.

\item The observed mass-metallicity relation at z$\sim$3 is difficult to reconcile with the simulations
by some hierarchical models, which predict metallicities higher than observed. If
aperture effects are taken into account then the discrepancy is probably even higher.
The main problem of these models seems to be that galaxies are assembled once they are already
evolved, from the stellar and chemical point of view.
Our results suggest that galaxies at z$>$3 are assembled mostly with relatively
un-evolved small galaxies, whose star formation efficiency is low. Most of the 
chemical evolution (hence most of the star formation) must occurs once small galaxies are
already assembled into bigger ones. This implies that most of the merging occurs before most of
the star formation.
Indeed, models and simulations
where a strong feedback keeps star formation low in the assembling small galaxies
(preventing them to evolve strongly from the stellar and chemical point of view),
provide a much better description of the mass-metallicity relation at z$\sim$3.5.

The finding that galaxies at z$>$3 are mostly assembled with un-evolved sub-units is not necessarily
in contrast with models of ``dry-merging'', i.e. models where galaxy assembly occurs through systems
that are significantly evolved and with little residual gas. Indeed, dry merging may be
the main mode of galaxy evolution at lower redshifts (z$<$3) without being in conflict
with our findings at z$>$3.

\end{enumerate}

\begin{acknowledgements}
We thank A.~Modigliani for his assistance in using the SINFONI pipeline.
We are grateful to C.~Kobayashi, S.~Savaglio and A.~Brooks for providing us the electronic
version of their simulations and data. We thank M.E.~De~Rossi for useful comments.
      Part of this work was supported by the Italian Institute for Astrophysics
	  (INAF) and by the Italian Space Agency (ASI) through contract
	  ASI-INAF I/016/07/0.
\end{acknowledgements}

\begin{table*}[!h]
\caption{Photometric properties and integration times
of the AMAZE sub-sample presented here}
\label{tab_samp}
{\centering
\begin{tabular}{lccccccc}
\hline\hline                 
Name & RA(J2000) & Dec(J2000) & z & $\rm R_{AB}$ & $\rm [3.6\mu m]_{AB}$ &
		$\rm [24\mu m]_{AB}$ & $\rm T_{int}$ \\
\hline
CDFa-C9 & 00 53 13.7 & +12 32 11.1 & 3.2119 &       23.99 &       22.66$^{}$ & $>$19.3 & 3.3 \\
CDFS-4414 & 03 32 23.2 & -27 51 57.9 & 3.4714 &       24.18 &       22.99$^{}$ & 20.7 & 6.7 \\
CDFS-4417 & 03 32 23.3 & -27 51 56.8 & 3.4733 &       23.42 &       22.23$^{}$ & $>$19.9 & 6.7 \\
CDFS-6664 & 03 32 33.3 & -27 50 07.4 & 3.7967 &       24.80 &       25.26$^{}$ & $>$19.9 & 5.0 \\
CDFS-16767 & 03 32 35.9 & -27 41 49.9 & 3.6241 &       24.13 &       24.04$^{}$ & $>$19.9 & 7.5 \\
CDFS-11991 & 03 32 42.4 & -27 45 51.6 & 3.6110 &       24.23 &       23.91$^{}$ & $>$19.9 & 7.5 \\
CDFS-2528 & 03 32 45.5 & -27 53 33.3 & 3.6877 &       24.64 &       24.17$^{}$ & $>$19.9 & 4.2 \\
SSA22a-M38 & 22 17 17.7 & +00 19 00.7 & 3.2940 &       24.11 &       21.75$^a$ & $>$19.6 & 6.7 \\
SSA22a-aug16M16 & 22 17 30.8 & +00 13 10.7 & 3.2920 &  23.83 &       22.87$^{}$ & $>$19.8 & 4.2 \\
\hline\hline                 
\end{tabular}
}
\\
The following quantities are reported in each column:
col. 1, object name; cols. 2-3, coordinates (J2000); col. 4, redshift spectroscopically
determined through the [OIII]5007 line in our spectra; col. 5, R-band AB-magnitude;
col. 6, Spitzer-IRAC 3.6$\mu$m AB magnitude; col. 7, Spitzer-MIPS 24$\mu$m AB magnitude
(lower limits are at 3$\sigma$);
col.8, on-source integration time (in unit of hours).\\
Note: $^a$ This is the AB magnitude at 4.5$\mu$m, since photometry at 3.6$\mu$m is not available.
\end{table*}

\begin{table*}[!h]
\caption{Physical properties of the sample inferred from their SED}
\label{tab_phys}
{\centering
\begin{tabular}{lccccccc}
\hline\hline                 
Name & $\rm log M_*$ (BC03) & $\rm log M_*$ (M05) & $\rm M_B$ & U--V & SFR & 
  E(B-V)$_*$  & age \\
     & $\rm [M_{\odot}]$    & $\rm [M_{\odot}]$   & [mag]     & [mag]& $\rm [M_{\odot} yr^{-1}]$
 & [mag]   & [Gyr] \\
\hline
CDFa-C9 & 10.18$\rm ^{+0.40}_{-0.08}$ &10.13$\rm ^{+0.30}_{-0.05}$ &-22.36 & -0.24 & 265 & 0.25$\rm ^{+0.05}_{-0.10}$ &0.06$\rm ^{+0.25}_{-0.01}$ \\
CDFS-4414 & 10.57$\rm ^{+0.19}_{-0.22}$ &10.48$\rm ^{+0.06}_{-0.23}$ &-22.47 & -0.02 & 113 & 0.20$\rm ^{+0.05}_{-0.10}$ &0.39$\rm ^{+0.60}_{-0.29}$ \\
CDFS-4417 & 10.29$\rm ^{+0.37}_{-0.11}$ &10.32$\rm ^{+0.29}_{-0.10}$ &-22.79 & -0.35 & 438 & 0.25$\rm ^{+0.05}_{-0.10}$ &0.05$\rm ^{+0.23}_{-9.99}$ \\
CDFS-6664 & 9.49$\rm ^{+0.10}_{-0.23}$ &9.30$\rm ^{+0.22}_{-0.08}$ &-20.93 & -0.70 & 35 & 0.10$\rm ^{+0.10}_{-0.04}$ &0.10$\rm ^{+0.09}_{-0.05}$ \\
CDFS-16767 & 10.05$\rm ^{+0.10}_{-0.16}$ &9.90$\rm ^{+0.13}_{-0.13}$ &-21.90 & -0.49 & 84 & 0.15$\rm ^{+0.05}_{-0.05}$ &0.12$\rm ^{+0.07}_{-0.07}$ \\
CDFS-11991 & 9.69$\rm ^{+0.12}_{-0.15}$ &9.64$\rm ^{+0.18}_{-0.08}$ &-21.64 & -0.46 & 55 & 0.10$\rm ^{+0.05}_{-0.04}$ &0.10$\rm ^{+0.09}_{-0.05}$ \\
CDFS-2528 & 9.76$\rm ^{+0.09}_{-0.07}$ &9.71$\rm ^{+0.16}_{-0.00}$ &-21.57 & -0.33 & 101 & 0.20$\rm ^{+0.05}_{-0.05}$ &0.06$\rm ^{+0.03}_{-0.01}$ \\
SSA22a-M38 & 11.01$\rm ^{+0.18}_{-0.41}$ &10.71$\rm ^{+0.19}_{-0.40}$ &-22.99 & 0.22 & 115 & 0.20$\rm ^{+0.05}_{-0.20}$ &0.63$\rm ^{+0.95}_{-0.45}$ \\
SSA22a-aug16M16 & 10.29$\rm ^{+0.20}_{-0.21}$ &10.15$\rm ^{+0.11}_{-0.28}$ &-22.32 & -0.23 & 42 & 0.06$\rm ^{+0.04}_{-0.06}$ &0.39$\rm ^{+0.39}_{-0.23}$ \\

\hline\hline                 
\end{tabular}
}
\\
The following quantities are reported in each column:
col. 1, object name;
col. 2, stellar mass inferred by using the galaxy templates of BC03;
col. 3, stellar mass inferred by using the galaxy templates of M05;
col. 4, absolute B magnitude;
col. 5, rest-frame U--V color;
col. 6, star formation rate (by using BC03);
col. 7, dust reddening affecting the stellar light, by using the attenuation curve of \cite{calzetti00};
col. 8, age of the stellar population.
\end{table*}

\begin{table*}[!h]
\caption{Line fluxes and metallicities inferred from the near-IR spectra}
\label{tab_met}
{\centering
\begin{tabular}{l|cccc|c}
\hline\hline                 
Name & F([OIII]5007) & F(H$\beta$) & F([OII]3727) & F([NeIII]3870) & 12+log(O/H) \\
     & \multicolumn{4}{c|}{$\rm 10^{-17}~erg~s^{-1}~cm^{-2}$$^a$} &  \\
\hline
CDFa-C9 & 6.83$\pm$0.21 & 1.44$\pm$0.18 & 2.16$\pm$0.14 & 0.53$\pm$0.13 & 8.10$\rm ^{+0.18}_{-0.21}$  \\
CDFS-4414 & 0.95$\pm$0.11 & 0.40$\pm$0.10 & 0.87$\pm$0.08 & $<$0.20 & 8.54$\rm ^{+0.15}_{-0.14}$  \\
CDFS-4417 & 2.00$\pm$0.11 & 0.89$\pm$0.09 & 1.17$\pm$0.11 & 0.20$\pm$0.09 & 8.55$\rm ^{+0.09}_{-0.10}$  \\
CDFS-6664 & 2.71$\pm$0.20 & 0.50$\pm$0.16 & 0.40$\pm$0.07 & --   & 7.73$\rm ^{+0.38}_{-0.29}$ \\
CDFS-16767 & 1.90$\pm$0.14 & 0.50$\pm$0.10 & 0.50$\pm$0.06 & --   & 8.31$\rm ^{+0.11}_{-0.17}$  \\
CDFS-11991 & 2.94$\pm$0.11 & 0.26$\pm$0.10 & 0.62$\pm$0.08 & 0.15$\pm$0.08 & 8.02$\rm ^{+0.17}_{-0.23}$  \\
CDFS-2528 & 1.87$\pm$0.26 & 0.32$\pm$0.18 & 0.67$\pm$0.16 & --   & 8.07$\rm ^{+0.39}_{-0.28}$  \\
SSA22a-M38 & 5.41$\pm$0.31 & 1.48$\pm$0.20 & 3.51$\pm$0.37 & $<$0.70 & 8.34$\rm ^{+0.15}_{-0.12}$  \\
SSA22a-aug16M16 & 1.65$\pm$0.10 & 0.30$\pm$0.07 & 0.45$\pm$0.11 & $<$0.25 & 7.99$\rm ^{+0.26}_{-0.34}$  \\
{\it Composite}$^a$ & 4.58$\pm$0.12 & 1.00$\pm$0.10 & 1.50$\pm$0.14 & 0.25$\pm$0.12 & 8.24$\rm ^{+0.15}_{-0.20}$ \\

\hline\hline                 
\end{tabular}
}
\\
The following quantities are reported in each column:
col. 1, object name; col. 2-5, emission line fluxes; col. 6, gas metallicity.\\
Notes:
$^a$ In the case of the composite spectrum fluxes are normalized
to the H$\beta$ flux (which is also subject to an error as listed in the corresponding column).
\end{table*}

\begin{table*}[!h]
\label{tab_met_cal}
\caption{Coefficients for different strong-line metallicity diagnostics 
in Eq.~\ref{eq_met_cal}.}
{\centering
\begin{tabular}{lccccc}
\hline\hline                 
Flux ratio ($\log{R}$) & $c_0$ & $c_1$ & $c_2$ & $c_3$ & $c_4$  \\
\hline
$\log{R_{23}}^a$                &  0.7462 & --0.7149 & --0.9401 & --0.6154 & --0.2524 \\
$\rm \log{[F([NII]6584)/F(H\alpha)]}$       & --0.7732 &  1.2357 & --0.2811 & --0.7201 & --0.3330 \\
$\rm \log{[F([OIII]5007)/F(H\beta)]}$       &  0.1549 & --1.5031 & --0.9790 & --0.0297 &  -- \\
$\rm \log{[F([OIII]5007)/F([OII]3727)]}$  & --0.2839 & --1.3881 & --0.3172 &  -- &  -- \\
$\rm \log{[F([OII]3727)/F(H\beta)]}$       &  0.5603 &  0.0450 & --1.8017 & --1.8434 & --0.6549 \\
$\rm \log{[F([OIII]5007)/F([NII]6584)]}$ &  0.4520 & --2.6096 & --0.7170 &  0.1347 &  -- \\
$\rm \log{[F([NeIII]3870)/F([OII]3727)]}$ & --1.2608 & --1.0861 & --0.1470 &  -- &  -- \\
\hline\hline                 
\end{tabular}
}
\\
Notes: $^a$ $\rm R_{23}=(F([OII]3727)+F([OIII]4959)+F([OIII]5007))/F(H\beta)$
\end{table*}

\begin{table*}[!h]
\caption{Best fit parameters for analytical form of the mass-metallicity relation
in Eq.~\ref{eq_mz_evol} at different redshifts}
\label{tab_mz_evol}
{\centering
\begin{tabular}{ccc}
\hline\hline                 
z & $\rm log M_0$ & $\rm K_0$ \\
\hline
0.07  &  11.18  & 9.04 \\
0.7  &  11.57   & 9.04 \\
2.2  &  12.38   & 8.99\\
3.5$^a$  &  12.76  & 8.79 \\
3.5$^b$  &  12.87  & 8.90 \\
\hline\hline                 
\end{tabular}
}
\\
Notes: $^a$ Value obtained by using the masses estimated with the
BC03 templates. $^b$ Value obtained by using the masses
estimated with the M05 templates.
\end{table*}

\end{document}